\documentstyle[epsfig]{mn}
\begin{document}

\title{Hyperluminous Infrared Galaxies}
\author[Rowan-Robinson]
{M. Rowan-Robinson 
\\
Astrophysics Group, Imperial College London, Blackett Laboratory,
Prince Consort Road, London SW7 2BZ;\\ 
}
\maketitle
\begin{abstract}
39 galaxies are now known, from follow-up of
faint IRAS sources and from submm observations of high redshift AGN, with
far infrared luminosities $ > 10^{13} L_{\odot}$.  13 of these, which have
been found in 60 or 850 $\mu$m surveys, form an important unbiased
sub-sample.  12 have been found by comparison of 60 $\mu$m surveys with
quasar or radio-galaxy catalogues, or from infrared surveys with colour selection
biased towards AGN, while a further 14 have been found through submm observations 
of known high redshift AGN.  
In this paper I argue, on the basis of detailed modelling of the spectral
energy distributions of hyperluminous galaxies with accurate radiative transfer models, 
and from evidence of high gas-mass in several cases, that the bulk of the emission from 
these galaxies at rest-frame wavelengths $\geq 50 \mu$m is due to
star formation.  Even after correction for the effects of lensing, hyperluminous galaxies 
with emission peaking at rest-frame wavelengths 
$\geq 50 \mu$m are therefore undergoing star-formation
at rates $> 10^{3} M_{\odot} yr^{-1}$ and are strong candidates for being
primeval galaxies, in the process of a major episode of star formation.

\end{abstract}
\begin{keywords}
infrared: galaxies - galaxies: evolution - star:formation - galaxies: starburst - cosmology: observations
\end{keywords}


\section{Introduction: Have we detected primeval galaxies ?}

This paper is directed towards the question; Have we already detected primeval galaxies ?
The characteristics of a primeval galaxy might be

\begin{itemize}
\item high redshift ($z > 1$)
\item undergoing a major episode of star formation  

(to form $10^{11} M_{\odot}$ in $\leq 10^{9}$ yrs, we need a star formation rate
$\dot{M}_{*} \geq 10^{2} M_{\odot} yr^{-1}$)

\item high gas fraction, say $M_{gas} \sim 10^{11} M_{\odot}$
\item evidence of interactions, mergers, dynamical youth.
\end{itemize}

First efforts to find such galaxies centred on spectroscopic searches for 
Ly$\alpha$-emitting galaxies (see eg the review by Djorgovski and Thompson 1992).  While 
examples of such galaxies are now being found, these early surveys suggested
that either star formation must be a more protracted process, occurring in smaller 
bursts (as expected in many bottom-up scenarios), or that dust extinction must play
a large role.

Steidel et al (1996) have shown that star-forming galaxies at z $>$ 3 can be found
through deep ground-based photometry in the U, G and R bands.  The high redshift galaxies manifest
themselves as U-band 'dropouts' as the Lyman limit absorption is redshifted into the U
band.  Over 500 spectroscopically confirmed high redshift galaxies have now been found
by this technique.  Many have weak or non-existent Lyman $\alpha$ emission, which accounts
for the lack of success of the spectroscopic surveys.  The role of dust in these galaxies
has been discussed by Pettini et al (1997, 1998a,b), Meurer et al (1997, 1998), Calzetti (1998). Pettini et al 
(1997, 1998a,b), Dickinson (1998) and Steidel et al (1998).  Even at redshift 3 it appears that dust
extinction may be appreciable.  However star formation rates in these galaxies are not
exceptional, typically 1-30 $M_{\odot}/yr$.
  
The first evidence for galaxies with very high rates of star formation came from
infrared surveys.  Starburst galaxies had been first identified by Weedman (1975) from their
ultraviolet excesses and characteristic emission-line spectra.  Prior to
the launch of IRAS, balloon and airborne measurements had demonstrated
that $L_{fir} > L_{opt}$ for several starburst galaxies (see review by Sanders and Mirabel 1996).
Joseph et al (1984) proposed that interactions and mergers might play
a role in triggering starbursts.
One of the major discoveries of the IRAS mission was the existence of ultraluminous 
infrared galaxies, 
galaxies with $L_{fir} > 10^{12} h_{50}^{-2} L_{\odot} (h_{50} = H_{o}/50)$.  
The peculiar Seyfert 2 galaxy Arp 220 was recognised as having an exceptional far infrared 
luminosity early in the mission (Soifer et al 1984).  

The conversion from far infrared luminosity to star formation rate has been discussed 
by many authors (eg Scoville and Young 1983, Thronson and Telesco 1986, Rowan-Robinson et al 1997).
Rowan-Robinson (1999) has given an updated estimate of how the star-formation
rate can be derived from the far infrared luminosity, finding

$\dot{M}_{*,all} /[L_{60}/L_{\odot}]$ = 2.2 $\phi/\epsilon$ x$10^{-10}$ 

where $\phi$ takes account of the uncertainty in the IMF (= 1, for a standard
Salpeter function) and $\epsilon$ is the fraction of uv light absorbed by dust, estimated
to by 2$/$3 for starburst galaxies (Calzetti 1998).
We see that the star-formation rates in
ultraluminous galaxies are $ > 10^{2} M_{\odot} yr^{-1}$.  However the time-scale
of luminous starbursts may be in the range $10^7 - 10^8$ yrs (Goldader et al 1997), so the
total mass of stars formed in the episode may typically be only 10$\%$ of the mass of a galaxy.

In this paper I discuss an even more extreme class of infrared galaxy, hyperluminous infrared
galaxies, which I define to be those with rest-frame infrared (1-1000 $\mu$m) luminosities, $L_{bol,ir}$,
in excess of $10^{13.22} h_{50}^{-2} L_{\odot}$ (=$10^{13.0} h_{65}^{-2} L_{\odot}$).  
For a galaxy with an M82-like starburst spectrum this corresponds to  
$L_{60} \geq 10^{13} h_{50}^{-2} L_{\odot}$, since the bolometric correction at 60 $\mu$m is 1.63.
  Sanders and Mirabel (1996) have a slightly
more stringent criterion, $L_{bol,ir} \geq 10^{13} h_{75}^{-2} L_{\odot}$, but in practice they 
use an estimate of $L_{bol}$ based on IRAS fluxes, which results in a demarcation almost identical
to that adopted here.  While the emission at
 rest-frame wavelengths 3-30 $\mu$m in these galaxies is often due to an AGN dust torus (see below),
 I argue that their emission at rest-frame wavelengths
$\geq 50 \mu$m is primarily due to extreme starbursts, implying star formation rates
in excess of 1000 $M_o/yr$.  These then are excellent candidates for being primeval galaxies, galaxies
undergoing a very major episode of star formation.

A preliminary version of these arguments was given by Rowan-Robinson (1996).  Granato et al (1996) 
modelled the seds of 4 hyperluminous galaxies (F10214, H1413, P09104 and F15307) in terms
of an AGN dust torus model.  Hughes et al (1997)
argued that hyperluminous galaxies can not be thought of as primeval galaxies on the basis of their
estimates of the gas mass and star formation rates in these galaxies.  However I show below that
their arguments are not compelling.  Fabian et al (1994, 1998)
argued from X-ray evidence that 09104+4109 and 15307+325 are obscured AGN and this 
interpretation is
supported for these objects by the non-detection of CO and submm continuum radiation 
(Yun and Scoville 1999, Evans et al 1999).  However the very severe upper limits on X-ray emission set by
Wilman et al (1999) for several hyperluminous infrared galaies led the latter to conclude
that the objects might be powered by starbursts.  McMahon et al (1999)
interpret the submillimetre emission from high redshift quasars and other hyperluminous infrared 
galaxies as powered, at least in part, by the AGN rather than by star formation.  
On the other hand Frayer et al (1999a,b)
favour a starburst interpretation for 14011+0252 and 02399-0136.
I discuss these arguments further below and try to resolve the question of what fraction of the
far infrared luminosity is powered by a starburst or AGN.  The use of model spectral energy
distributions derived from accurate radiative transfer codes is a significant advance on some
previous work.

I assume throughout that $H_o$ = 50, $\Omega_o$ = 1.  With lower $\Omega_o$, more galaxies,
especially at higher redshifts, would satisfy the definition adopted.

\section{Properties of ultraluminous infrared galaxies}
Sanders et al (1988) discussed the properties of 10 
IRAS ultraluminous galaxies with 60 $\mu$m fluxes $>$ 5 Jy and concluded
that (a) all were interacting, merging or had peculiar morphologies, (b) all 
had AGN line spectra.  On the other hand Leech et al (1989) found that only  2 of their 
sample of 6 ultraluminous 
IRAS galaxies had an AGN line spectrum.  Leech et al (1994) found that 67 
$\%$ of a much larger sample (42) of 
ultraluminous galaxies were interacting, merging or peculiar.  Lawrence et al 1989 had 
found a much lower fraction 
amongst galaxies of high but less extreme infrared luminosity.  The 
incidence of interacting, merging or peculiar galaxies by ir luminosity is summarised in 
Fig 1 of Rowan-Robinson (1991): the proportion of galaxies which are peculiar, interacting
or merging increases steadily from 10-20$\%$ at low ir luminosities to $> 80\%$ for
ultraluminous ir galaxies.  
The situation on point (b) remains controversial, though, since Lawrence et al (1999) 
find only a fraction
21$\%$ of 81 ultraluminous galaxies in the QDOT sample to have AGN spectra (but on the basis
only of low-resolution spectra).  Veilleux et al (1995)
find for a smaller sample (21 galaxies) that 33$\%$ of ultraluminous galaxies are Seyfert 1 or 2, 
with an additional 29$\%$ having liner spectra, which they also classify as AGN.  Veilleux et al
(1999a) find that 24$\%$ of 77 galaxies with $10^{12} < L_{ir} < 10^{12.3}$ are Seyfert 1 or 2,
increasing to 49$\%$ of 31 galaxies with $10^{12.3} < L_{ir} < 10^{12.8}$ (for $H_o = 75$).  Veilleux et al
(1999b) find, from near-ir imaging studies, no evidence that liners should be considered to be AGN.
Sanders (1999) reports that most nearby ultraluminous ir galaxies contain an AGN at some level.

Rowan-Robinson and Crawford (1989) found that their standard starburst galaxy model gave 
an excellent fit to the far 
infrared spectrum of Mk 231, an archetypal ultraluminous ir galaxy.  However their models 
for Arp 220 appeared to 
require a much higher optical depth in dust than the typical starburst galaxy.  Condon et 
al (1991) showed that 
the radio properties of most ultraluminous ir galaxies were consistent with a starburst 
model and argued that many of these 
galaxies required an exceptionally high optical depth.  This suggestion was confirmed by 
the detailed models of 
Rowan-Robinson and Efstathiou (1993) for the far infrared spectra of the Condon et al 
sample.

Quasars and Seyfert galaxies, on the other hand, tend to show a characteristic mid 
infrared continuum, broadly flat 
in $\nu S_{\nu}$ from $3-30 \mu$m.  This component was modelled by 
Rowan-Robinson and Crawford (1989) as dust in the 
narrow-line region of the AGN with a density distribution n(r) $\alpha$  $r^{-1}$ .  
More realistic models of this 
component based on a toroidal geometry are given by Pier and Krolik 
(1992), Granato and Danese (1994), Rowan-Robinson (1995), Efstathiou and 
Rowan-Robinson (1995).  Rowan-Robinson 
(1995) suggested that most quasars contain both (far ir) starbursts and (mid ir) 
components due to (toroidal) dust 
in the narrow line region.

Rowan-Robinson and Crawford (1989) were able to fit the IRAS colours and spectral
energy distributions of galaxies detected in all 4 IRAS bands with a mixture of 3 components,
emission from interstellar dust ('cirrus'), a starburst and an AGN dust torus.  Recently
Xu et al (1998) have shown that the same 3-component approach can be used to fit
the ISO-SWS spectra of a large sample of galaxies.  To accomodate the Condon et al (1991)
and Rowan-Robinson and Efstathiou (1993) evidence for higher optical depth starbursts, 
Ferrigno et al (1999) have extended the Rowan-Robinson and Crawford (1989) analysis
to include a fourth component, an Arp220-like, high optical depth starburst, for
galaxies with log $L_{60} >$ 12.  Efstathiou et al (1999) have given improved
radiative transfer models for starbursts as a function of the age of the starburst,
for a range of initial dust optical depths.

Sanders et al (1988) proposed, on the basis of spectroscopic arguments for a sample of 10 objects,
that all ultraluminous infrared galaxies contain an AGN and that the far infrared emission is
powered by this.  Sanders et al (1989) proposed a specific model, in the context of
a discussion of the infrared emission from PG quasars, that 
the far infrared emission comes from the outer parts of
a warped disk surrounding the AGN.  This is a difficult hypothesis to disprove,
because if an arbitrary density distribution of dust is allowed at large distances from the
AGN, then any far infrared spectral energy distribution could in fact be generated.
In this paper I consider whether the AGN dust torus model of Rowan-Robinson (1995) 
can be extended naturally to explain the far infrared and submilllimetre emission
from hyperluminous infrared galaxies, but conclude that in many cases this does
not give a satisfactory explanation.  I also place considerable weight on
whether molecular gas is detected in the objects via CO lines.

Rigopoulou et al (1996) observed a sample of ultraluminous infrared galaxies from
the IRAS 5 Jy sample at submillimetre wavelengths, with the JCMT, and at X-ray wavelengths, 
with ROSAT.  They found that most of the far infrared and submillimetre spectra were fitted well
with the starburst model of Rowan-Robinson and Efstathiou (1993).  The ratio of bolometric
luminosities at 1 keV and 60 $\mu$m lie in the range $10^{-5} - 10^{-4}$ and are
consistent with a starburst interpretation of the X-ray emission in almost all cases.
Even more conclusively, Lutz et al (1996) and Genzel et al (1998) have used ISO-SWS spectroscopy to show that 
the majority of ultraluminous ir galaxies are powered by a starburst rather than an AGN.

\section{Hyperluminous Infrared Galaxies}

In 1988 Kleinmann et al identified P09104+4109 with a z = 0.44 galaxy, implying a total 
far infrared luminosity of 
1.5x$10^{13}$, a factor 3 higher than any other ultraluminous galaxy seen to that date.  
In 1991, as part of a 
programme of systematic identification and spectroscopy of a sample of 
3400 IRAS Faint Source Survey (FSS) sources, Rowan-Robinson et al discovered IRAS F10214+4724, 
an IRAS galaxy with z = 2.286 and a far 
infrared luminosity of 3x$10^{14} h_{50}^{-2} L_{\odot}$ .  This object 
appeared to presage an entirely new class of infrared galaxies.  The detection of a huge 
mass of CO by Brown and vandenBout (1991), $10^{11} h_{50}^{-2} M_{\odot}$, confirmed by the 
detection 
of a wealth of molecular lines (Solomon et al 1992), and of submillimetre 
emission at wavelengths 450-1250 $\mu$m (Rowan-Robinson et al 1991,1993, Downes 
et al 1992), implying a huge mass of dust, $10^{9} h_{50}^{-2} M_{\odot}$  
confirmed that this was an exceptional object.  Early models suggested this might be 
a giant elliptical galaxy 
in the process of formation (Elbaz et al 92).  Simultaneously with the growing evidence 
for an exceptional starburst 
in F10214, the Seyfert 2 nature of the emission line spectrum 
(Rowan-Robinson et al 1991, Elston et al 1994a) was supported by the evidence for very 
strong optical polarisation 
(Lawrence et al 93, Elston et al 94b).  Subsequently it has become clear that F10214 
is a gravitationally lensed system 
(Graham and Liu 1995, Broadhurst and Lehar 1995, Serjeant et al 1995, Eisenhardt et al 1996)
 with a magnification 
of about 10 at far infrared wavelengths, but not much greater than that (Green and 
Rowan-Robinson 1996).  Even when 
the magnification of 10 is allowed for, F10214 is still an exceptionally luminous far ir source.

In 1992 Barvainis et al successfully detected submillimetre emission from the  z=2.546 
'clover-leaf' gravitationally 
lensed QSO, H1413+117, which suggested that H1143 is of similar luminosity to F10214.
Subsequently (Barvainis et al 1995) they realized that the galaxy was an IRAS FSC source.

Here I want to place emphasis on the hyperluminous galaxies detected as a result of
unbiassed surveys at far infrared (and submillimetre) wavelengths.
The program of follow-up of IRAS FSS sources which led to the discovery of F10214 
(Rowan-Robinson et al 1991, Oliver et al 1996) has also resulted in the discovery 
of a further 6 galaxies or quasars with far ir luminosities $> 10^{13.22} h_{50}^{-2} L_{\odot}$ 
(McMahon et al 99).  Four galaxies from the PSCz survey (Saunders et al 1996) of IRAS galaxies 
brighter than  0.6 Jy at 60 $\mu$m fall into the hyperluminous category (a further two are blazars,
3C345 and 3C446, and these are not considered further here), and a further example has been
found by Stanford et al (1999) in a survey based on a comparison of the IRAS FSS with the VLA
FIRST radio survey.  Two galaxies detected in submillimetre surveys at 850 $\mu$m with SCUBA
also fall into the hyperluminous category (but one of these only because of the effect of 
gravitational lensing).

Cutri et al (1994) reported a search for IRAS FSS 
galaxies with 'warm' 25/60 $\mu$m colours, which yielded the 
z = 0.93 Seyfert 2 galaxy, F15307+3252 (see also Hines et al 1995).  Wilman et al (1999)
have reported a further 2 hyperluminous galaxies from a more recent search for 'warm'
galaxies by Cutri et al (1999, in preparation).
Dey and van Breugel (1995) 
reported a comparison of the Texas radio survey 
with the IRAS FSS catalogue, which resulted in 5 galaxies with fir ir 
luminosities $> 10^{13} h_{50}^{-2} L_{\odot}$.  However three of these are present only in the
FSS Reject Catalogue and have not been able to be confirmed as far infrared sources to date.  The
other two are discussed below.  Four PG quasars from the list of Sanders et al (1989)
(two of which are part of the study of Rowan-Robinson (1995)), fall
into the hyperluminous category.  Recently Irwin et al (1999) have found a z = 3.91
quasar which is associated with with IRAS FSS source F08279+5255, the highest redshift IRAS object
to date.    

Finally, inspired by the success in finding highly redshifted submillimetre continuum and 
molecular line emission 
in F10214, several groups have studied an ad hoc selection of very high redshift quasars 
and radio-galaxies, with 
several notable successes (Andreani et al 1993, Dunlop et al 1995, Isaak et al 1995, McMahon
 et al 1995b, Ojik et al 1995,  Ivison 1995, Omont et al 1997, Hughes et al 1997, McMahon et al 1999).  
Many of these detections 
imply far ir luminosities $> 10^{13.22} h_{50}^{-2} L_{\odot}$ , assuming that 
the far ir spectra are typical starbursts.  
  In all there are now
39 hyperluminous infrared galaxies known, which are listed in Tables 1-3 according to whether they are
(1) the result of unbiased 60 $\mu$m (or submm) surveys, (2) found from comparison of known quasar and 
radio-galaxy lists with 60 $\mu$m catalogues, (3) found through submillimetre observations of
known high redshift AGN.  Table 4 lists some luminous infrared galaxies which do not quite
meet my criteria, but have far infrared luminosities $>10^{13.0} L_{\odot}$.  But to set these in perspective
there are a further 20 PSCz galaxies which have 13.00 $< log_{10} L_{ir}/L_{\odot} <$ 13.22
 (for $H_o$ = 50).  

From the surveys summarised in Table 1 we can estimate that the number of hyperluminous galaxies per
sq deg brighter than 200 mJy at 60 $\mu$m is 0.0027-0.0043, which would imply that there
are 100-200 hyperluminous IRAS galaxies over the whole sky, 25 of which are listed in Tables
1 and 2.

\section{Models for Hyperluminous Infrared Galaxies}

For a small number of these galaxies we have reasonably detailed continuum spectra from 
radio to uv wavelengths.  
Figures 1-17 show the infrared continua of these hyperluminous galaxies, with fits using radiative 
transfer models (specifically the standard M82-like starburst model and
an Arp220-like high optical depth starburst model from Efstathiou et al 
(1999) and the standard AGN dust torus model of Rowan-Robinson (1995).  More than half of those
shown have measurements at at least 9 independent wavelengths

We now discuss the individual objects (and a few not plotted) in turn:

$\bf{F10214+4724}$

The continuum emission from F10214 was the subject of a detailed discussion by Rowan-Robinson 
et al (1993).  
Green and Rowan-Robinson (1995) have discussed starburst and AGN dust tori models for F10214.  
Fig 1 shows M82-like and Arp 220-like starburst
models fitted to the sumbillimetre data for this galaxy.  The former gives a
good fit to the latest data.  The 60 $\mu$m flux requires an AGN dust torus
component.  To accomodate the upper limits at 10 and 20 $\mu$m, it is necessary
to modify the Rowan-Robinson (1995) AGN dust torus model so that the maximum temperature
of the dust is 1000 K rather than 1600 K.  I have also shown the effect of allowing the
dust torus to extend a further factor 3.5 in radius.  This still does not account for the amplitude
of the submm emission.  The implied extent of the narrow-line region for this extended AGN
dust torus model, which we 
use for several other objects, would be 326 $(L_{bol}/10^{13} L_{\odot})^{1/2}$ pc 
consistent with 60-600 $(L_{bol}/10^{13} L_{\odot})^{1/2}$ pc quoted by Netzer (1990).
  Evidence for a strong starburst contribution to the ir emission from F10214
is given by Kroker et al (1996) and is supported by the high gas mass detected via
CO lines (see section 6).  Granato et al (1996) attempt to model the whole sed of F10214
with an AGN dust torus model, but still do not appear to be able to account for the 60 $\mu$m
emission.

$\bf{F0023+1024}$

A starburst model fits the IRAS and ISO data (Verma et al 1999) well and there is a strong limit
on any AGN dust torus contribution
.

$\bf{SMMJ02399-0136}$

A starburst model fits the submm data well and the ISO detection at 15 $\mu$m
gives a very severe constraint on any AGN dust torus component.  The starburst 
interpretation of the submm emission is supported by the gas mass estimated from CO detections
(Frayer et al 1999b).

$\bf{F1421+3845}$

A starburst model is the most likely interpretation of the IRAS and ISO 60-180
$\mu$m data, but there are discrepancies.  There is a strong limit
on any AGN dust torus component
.

$\bf{TX0052+4710}$

There is little evidence for a starburst contribution to the sed of this galaxy.
An extended AGN dust torus model fits the data reasonably well, apart from the ISO detection
by Verma et al (1999) at 180 $\mu$m.

$\bf{F08279+5255}$

An M82-like starburst is a good fit to the submm data and an AGN dust torus model is a good
fit to the 12-100 $\mu$m data.  The high gas mass detected
via CO lines (Downes et al 1998) supports a starburst interpretation, though the ratio of 
$L_{sb}/M_{gas}$ is on the high side (see section 6).  The submm data can also
be modelled by an extension of the outer radius of the AGN dust torus and in this case the
starburst luminosities given in Table 2 will be upper limits.

$\bf{P09104+4109}$

The 100 $\mu$m upper limit places a limit on the starburst contribution and there is
no detection of CO emission in this galaxy.  The 12-60 $\mu$m data can be fitted
by the extended AGN dust torus model. 

$\bf{PG1148+549}$

A starburst is a natural explanation of the 100 $\mu$m excess compared to the AGN dust torus
fit to the 25 and 60 $\mu$m data, but detection in the submm and in CO would be important
for confirming this interpretation.

$\bf{PG1206+459}$

The IRAS 12-60 $\mu$m data and the ISO 12-200 $\mu$m data (Haas et al 1998) are well-fitted 
by the extended AGN dust torus model and there is no evidence for a starburst.

$\bf{H1413+117}$

The submm data is well fitted by an M82-like starburst and the gas mass implied by the 
CO detections (Barvainis et al 1994) supports this interpretation.  The extended AGN dust torus model
discussed above does not account for the submm emission.  However Granato et al
(1996) model the whole sed of H1413 in terms of an AGN dust torus model.

$\bf{15307+325}$

A starburst model gives a natural explanation for the 60-180 $\mu$m excess compared
to the AGN dust torus model required for the 6.7 and 14.3 $\mu$m emission (Verma et al 
1999), but the non-detection of CO poses a problem for a starburst intepretation.

$\bf{PG1634+706}$

The IRAS 12-100 $\mu$m data and the ISO 150-200 $\mu$m data (Haas et al 1998) are well-fitted 
by the extended AGN dust torus model
and an upper limit can be placed on any starburst component.  The non-detection of CO is
consistent with this upper limit.

$\bf{4C0647+4134}$

An M82-like starburst model gives  a reasonable fit to the submm data (although
the 1250 $\mu$m flux seems very weak).  Observations
in the mid-ir are needed to constrain the AGN dust torus.  Since the AGN is not
seen directly we have no constraints on its optical luminosity.  We have used the
non-detection of this source by IRAS (which we take to imply S(60) $<$ 250 mJy)
to set a limit on $L_{tor}$.  This limit is not strong enough to rule out an
AGN dust torus interpretation of the submm data.

$\bf{BR1202-0725}$

The submm data are well-fitted by an M82-like starburst model.  The QSO is
seen directly, although with strong self-absorption in the lines (Storrie-
Lombardi et al 1996), so we can probably not use the QSO bolometric
luminosity to set a limit on $L_{tor}$.    
Wilkes et al (1998) have detected this galaxy at 7, 12 and 25 $\mu$m
with ISO, which would imply that the QSO is undergoing very strong extinction.    
An AGN dust torus model is capable of accounting for the whole spectrum but the starburst 
interpretation of the submm emission is supported by the gas mass estimated from CO detections
(Ohta et al 1996, Omont et al 1996).

$\bf{PG1241+176}$

The ISO data of Haas et al (1999) can be fitted with an AGN dust torus model and
there is no evidence for a starburst component.  The 1.3 mm flux is probably an
extrapolation of the radio continuum.

$\bf{PG1247+267}$
The ISO data of Haas et al (1999) can be fitted with an AGN dust torus model and
there is no evidence for a starburst component.

$\bf{PG1254+047}$
The ISO data of Haas et al (1999) can be fitted with an AGN dust torus model and
there is only weak evidence for a starburst component.

$\bf{BRI1335-0417}$

The submm data can be fitted with a starburst model and since the QSO
is seen directly we can use its estimated bolometric luminosity to set a limit
on $L_{tor}$, which makes it unlikely that the submm emission is from an
dust torus.    The starburst 
interpretation is supported by gas mass estimated from CO detections
(Guilloteau et al 1997).

$\bf{PC2047+0123}$

The limit on $L_{tor}$ from the bolometric luminosity of the QSO makes it unlikely
that  an AGN dust torus is responsible for the 350 $\mu$m emission.  However
a starburst model can not fit both the 350 and 1250 $\mu$m observed fluxes.
Observations at other submm wavelengths may help to clarify the situation.

\medskip

For the remaining objects we have only 60 $\mu$m or single submillimetre 
detections and for these we estimate their far infrared luminosity, and other properties,
using the standard 
starburst model of Efstathiou et al (1999).  Tables 1-4 give the luminosities inferred in
the starburst ($L_{sb}$) (and fits of the A220 model in brackets) and AGN dust torus ($L_{tor}$) 
components, or limits on these,
with an indication, from the row position of the estimate, of which wavelength the estimate
is made at.  In Fig 18 we show the far infrared 
luminosity derived for an assumed starburst component, versus 
redshift, for hyperluminous galaxies, with lines indicating observational 
constraints at 60, 800 and 1250 $\mu$m.  Three of the sources with 
(uncorrected) total bolometric luminosities 
above $10^{14} h_{50}^{-2} L_{o}$ are strongly gravitationally 
lensed.  IRAS F10214+4724
was found to be lensed with a magnification which ranges from 100 at optical wavelengths
to 10 at far infrared wavelengths (Eisenhardt et al 1996, Green and Rowan-Robinson 1996).
The 'clover-leaf' lensed system H1413+117 has been found to have a magnification of
10 (Yun et al 1997).  Downes et al (1998) report a magnification of 14 for F08279+5255.
Also, Ivison et al estimate a magnification of 2.5 for SMMJ02399-0136 and Frayer et al (1999a)
quote a magnification of 2.75 $\pm$0.25 for SMMJ14011+0252.

These magnifications have to be corrected for in estimating luminosities (and dust and
gas masses) and these corrections are indicated in Fig 18.  It appears to be
a reasonable assumption that if a starburst luminosity in excess of $10^{14} L_{\odot}$
is measured then the source is likely to be lensed, so F14218 and TX1011 merit further
more careful study.  


\begin{figure}
\epsfig{file=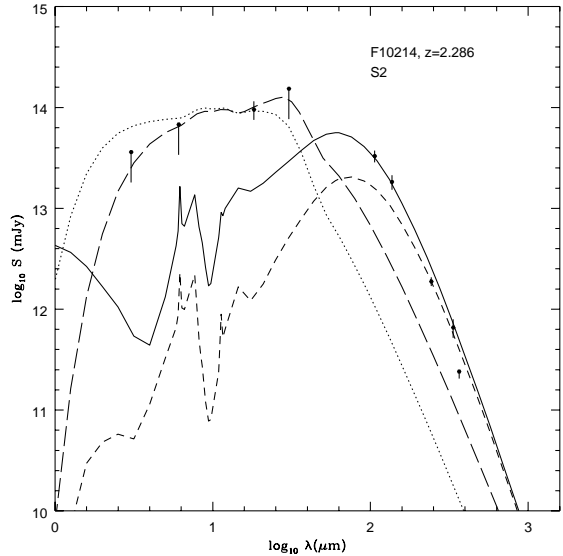,angle=0,width=8cm}
\caption{
Observed spectral energy distribution for F10214, modelled with M82-type starburst
(solid curve), Arp 220-type starburst (broken curve), AGN dust torus (dotted curve, modified AGN
dust torus model - long-dashed curve).}
\end{figure}

\begin{figure}
\epsfig{file=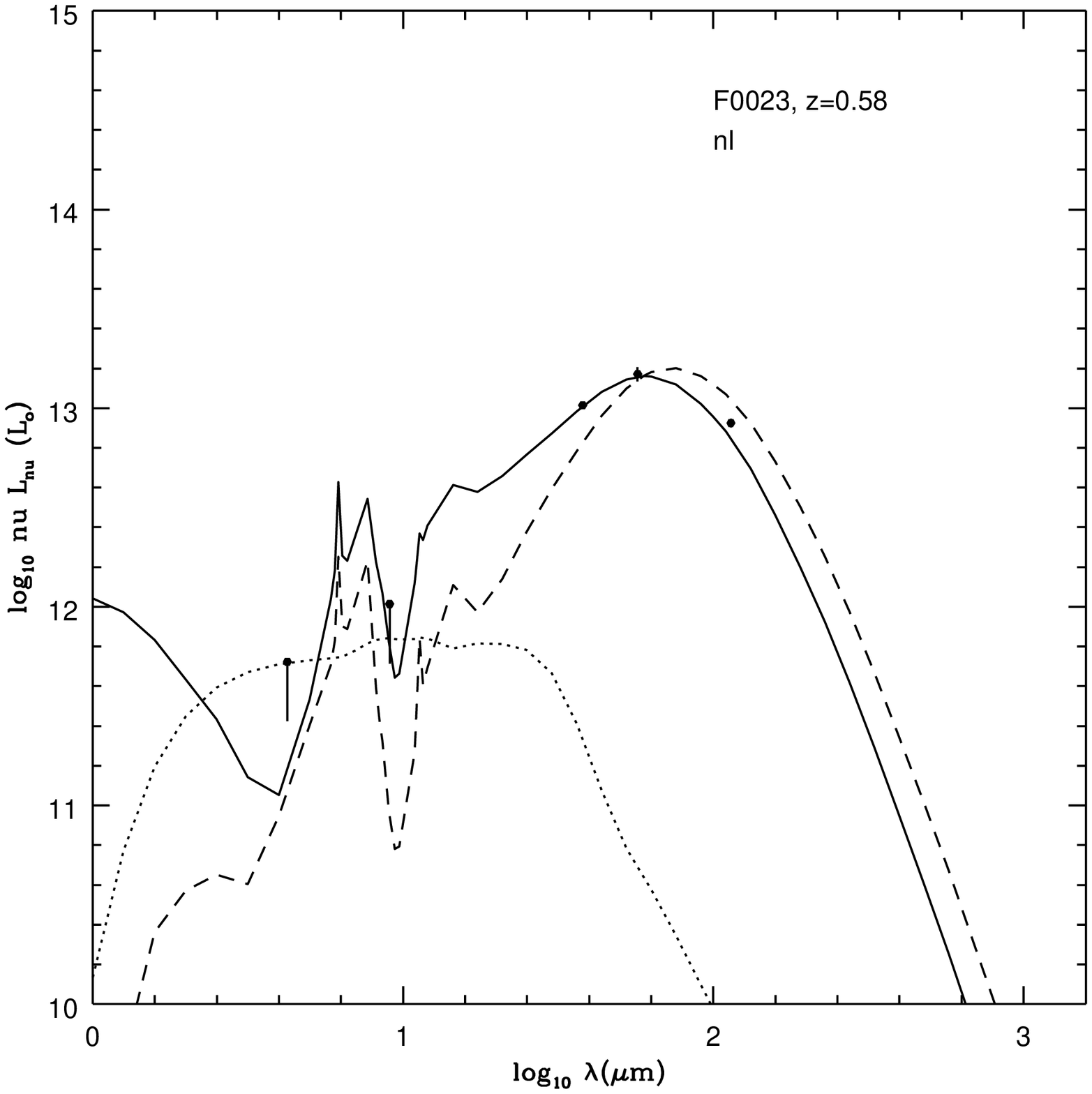,angle=0,width=8cm}
\caption{
Observed spectral energy distribution for F0023, notation as for Fig 1.}
\end{figure}

\begin{figure}
\epsfig{file=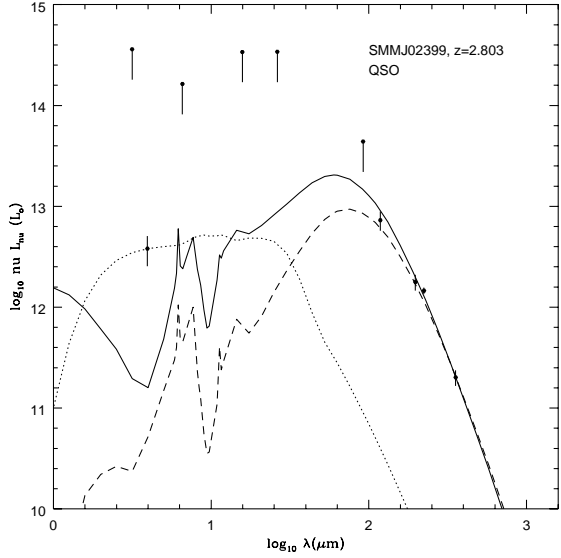,angle=0,width=8cm}
\caption{
Observed spectral energy distribution for SMMJ02399, notation as for Fig 1.}
\end{figure}

\begin{figure}
\epsfig{file=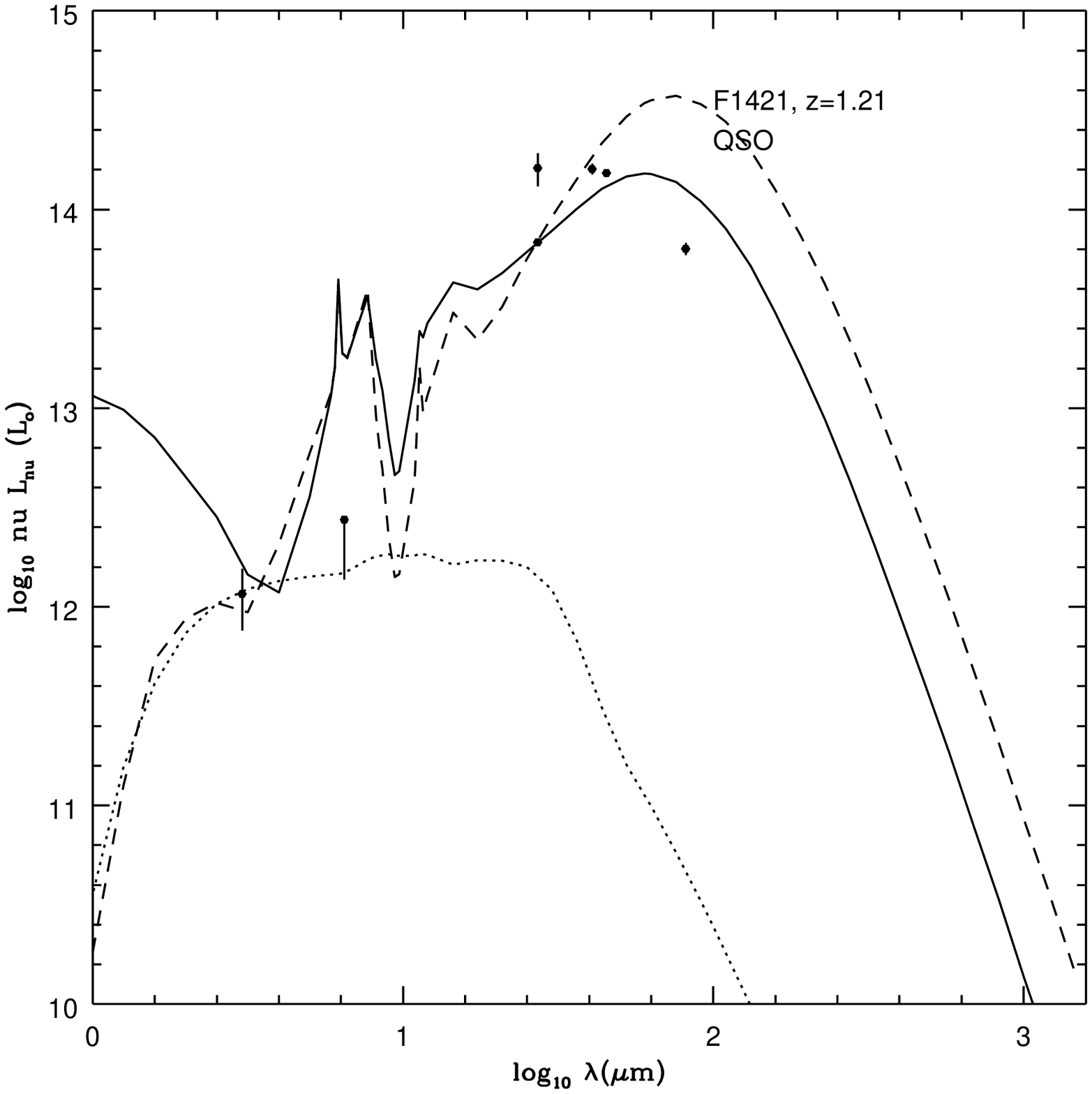,angle=0,width=8cm}
\caption{
Observed spectral energy distribution for F1421, notation as for Fig 1.}
\end{figure}

\begin{figure}
\epsfig{file=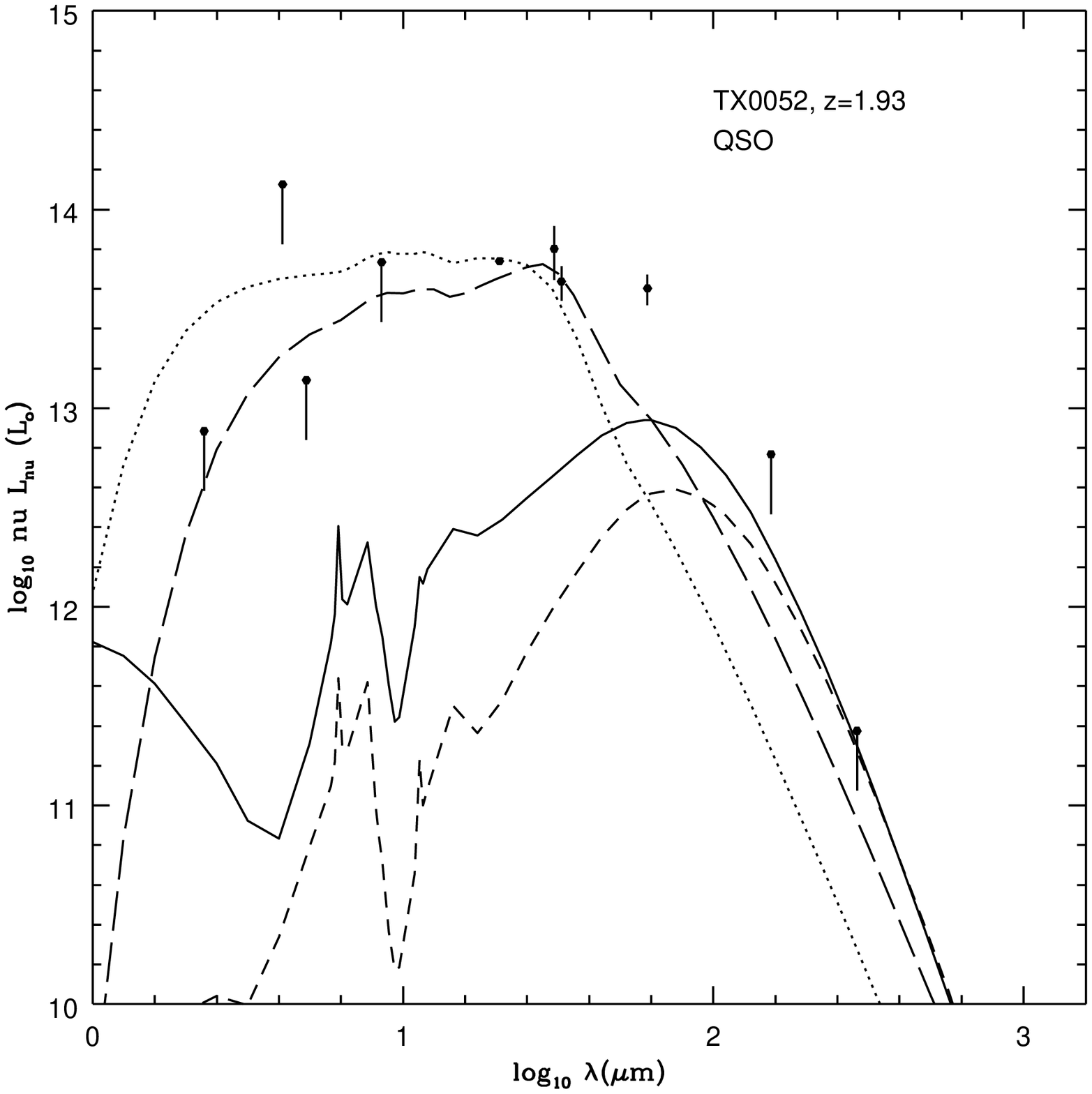,angle=0,width=8cm}
\caption{
Observed spectral energy distribution for TX0052, notation as for Fig 1.}
\end{figure}

\begin{figure}
\epsfig{file=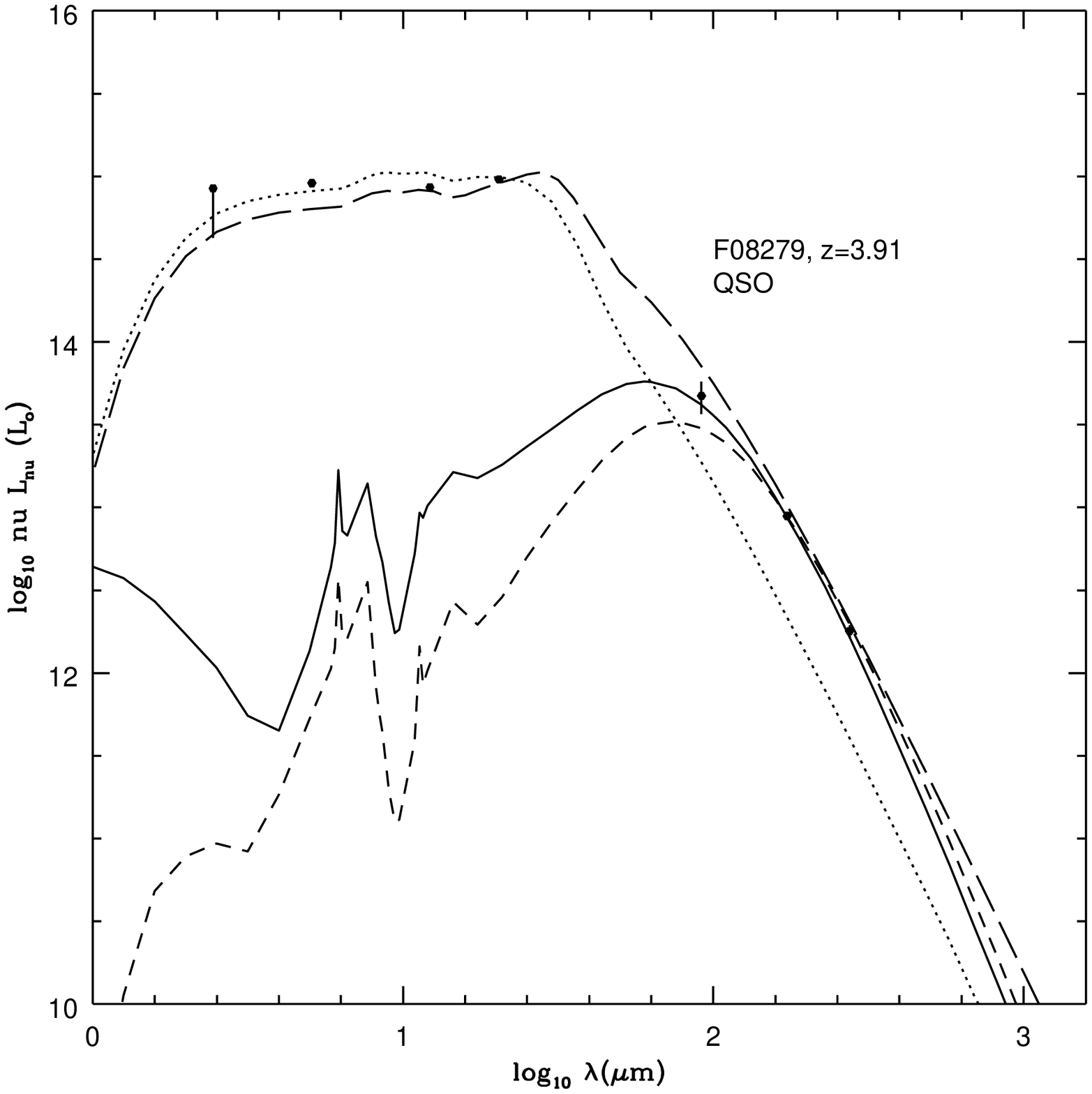,angle=0,width=8cm}
\caption{
Observed spectral energy distribution for F08279, notation as for Fig 1.}
\end{figure}

\begin{figure}
\epsfig{file=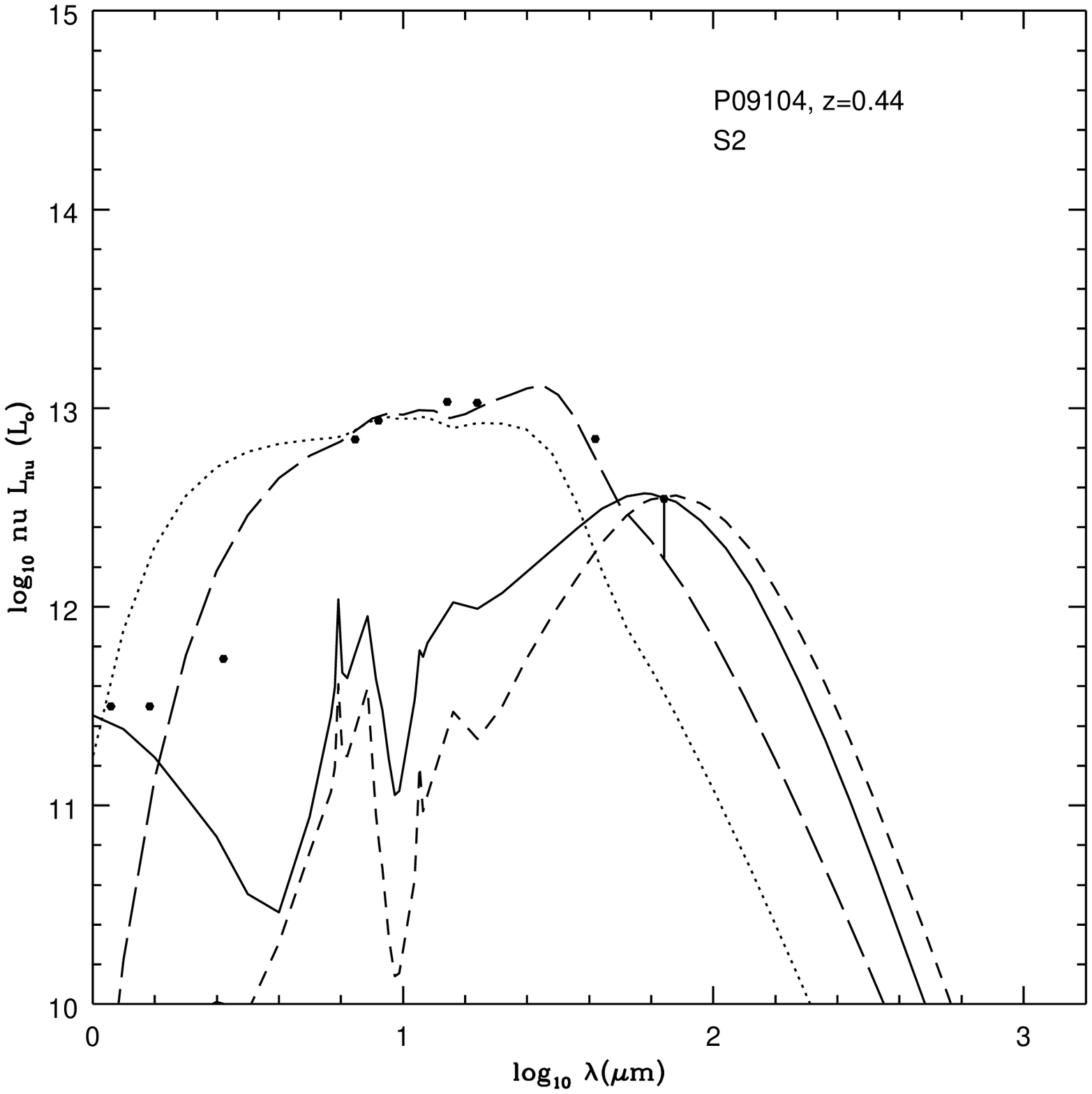,angle=0,width=8cm}
\caption{
Observed spectral energy distribution for P019104, notation as for Fig 1.}
\end{figure}

\begin{figure}
\epsfig{file=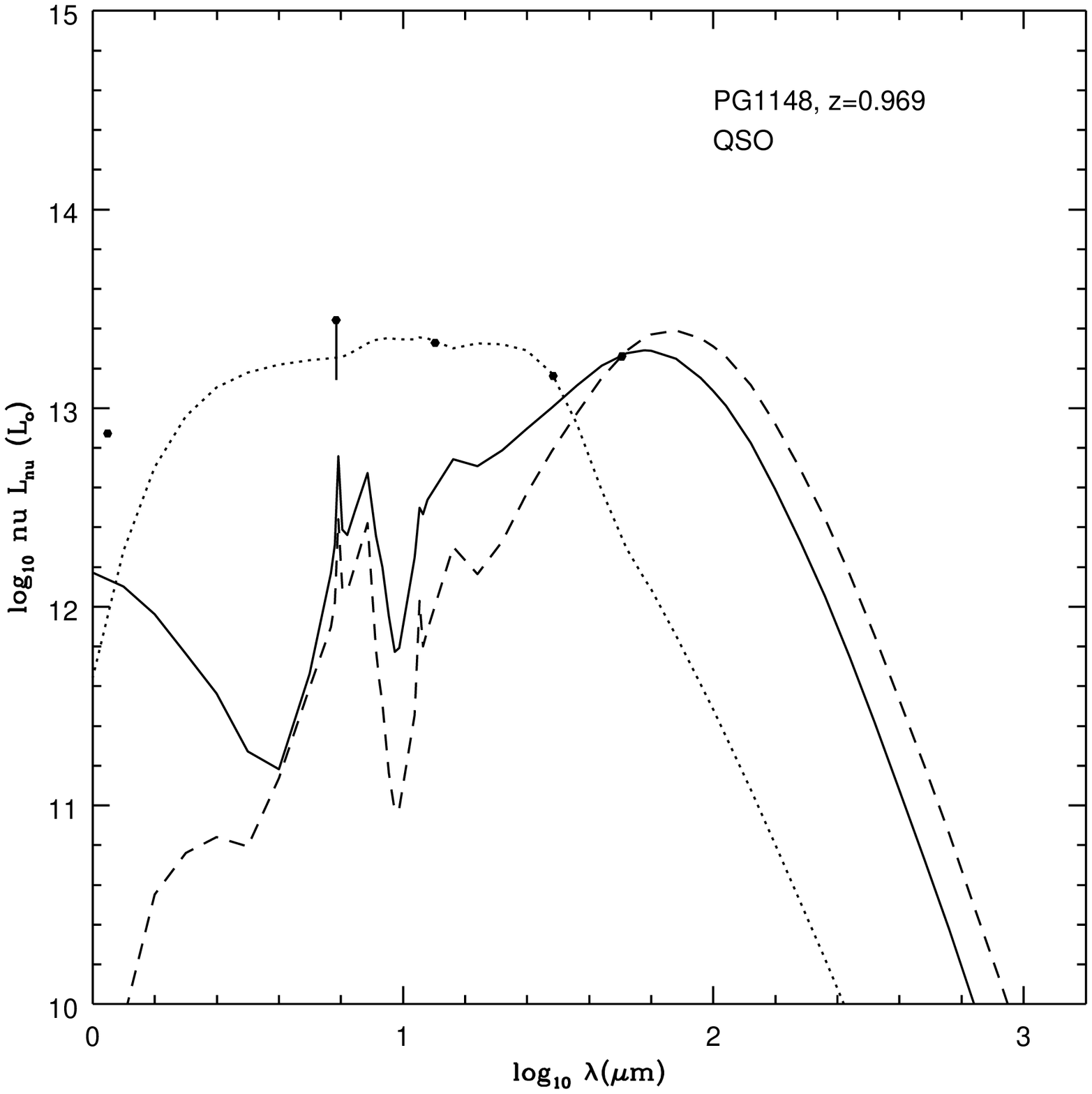,angle=0,width=8cm}
\caption{
Observed spectral energy distribution for PG1148, notation as for Fig 1.}
\end{figure}

\begin{figure}
\epsfig{file=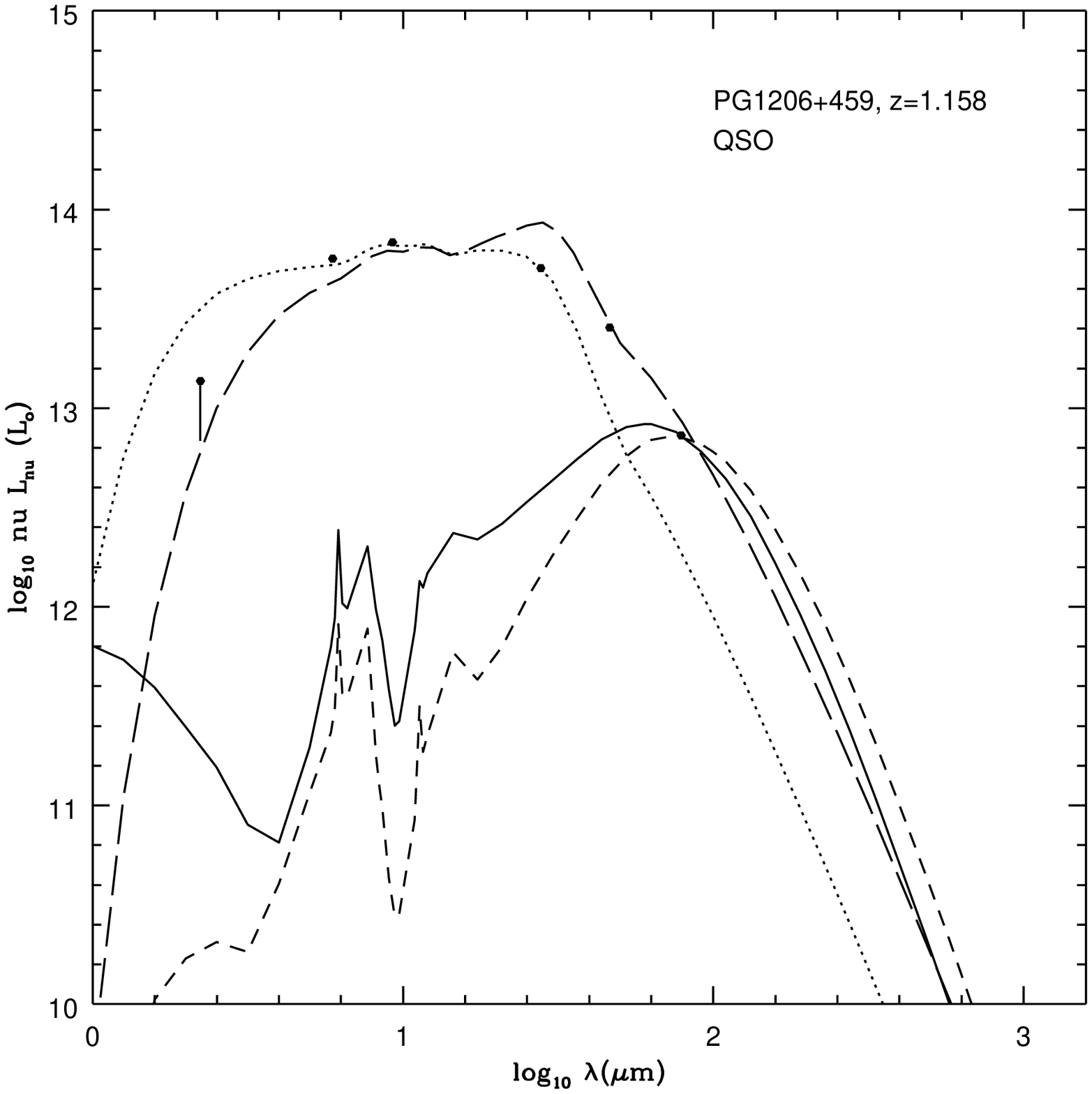,angle=0,width=8cm}
\caption{
Observed spectral energy distribution for PG1206, notation as for Fig 1.}
\end{figure}

\begin{figure}
\epsfig{file=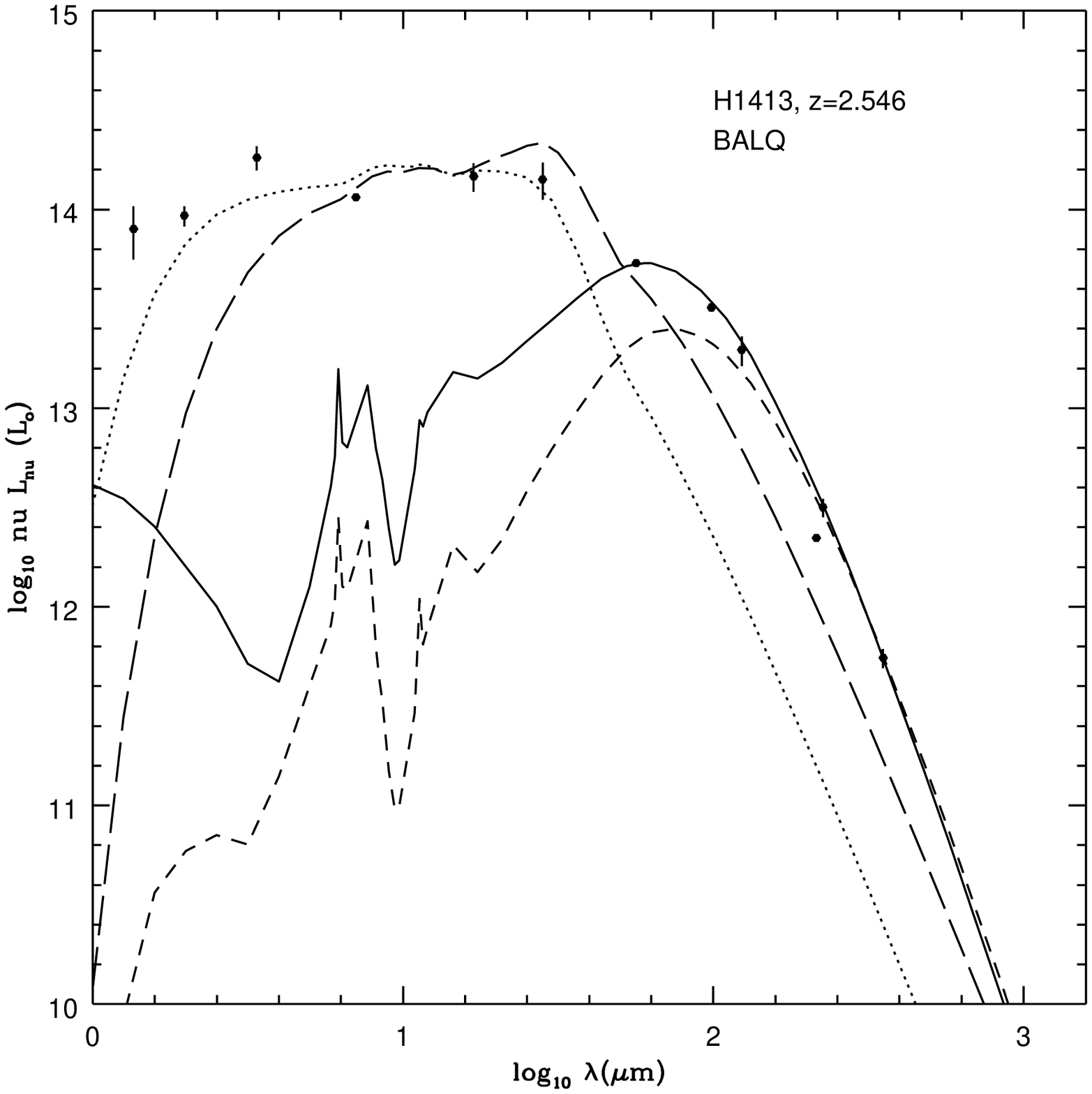,angle=0,width=8cm}
\caption{
Observed spectral energy distribution for H1413, notation as for Fig 1.}
\end{figure}

\begin{figure}
\epsfig{file=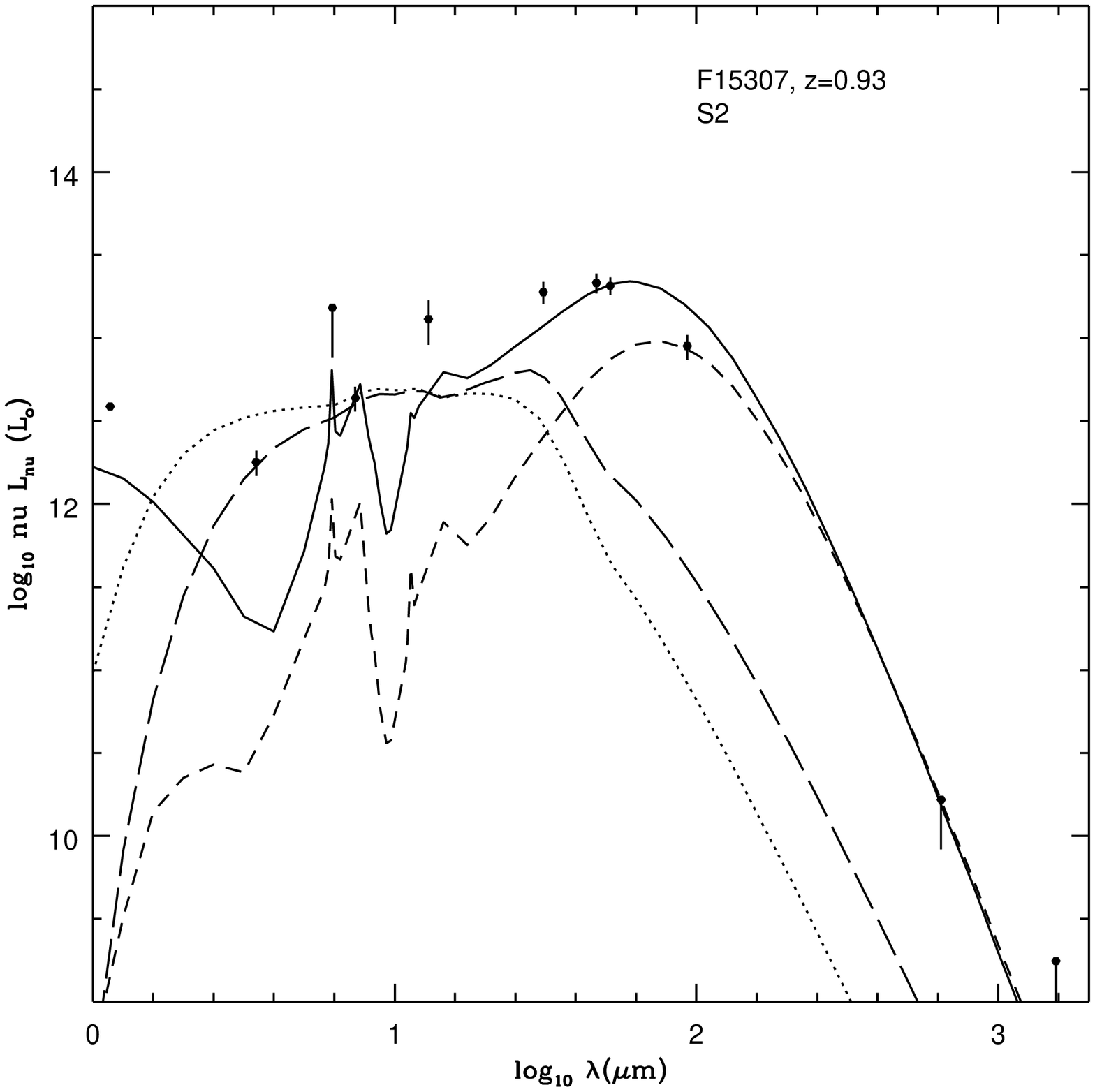,angle=0,width=8cm}
\caption{
Observed spectral energy distribution for F15307, notation as for Fig 1.}
\end{figure}

\begin{figure}
\epsfig{file=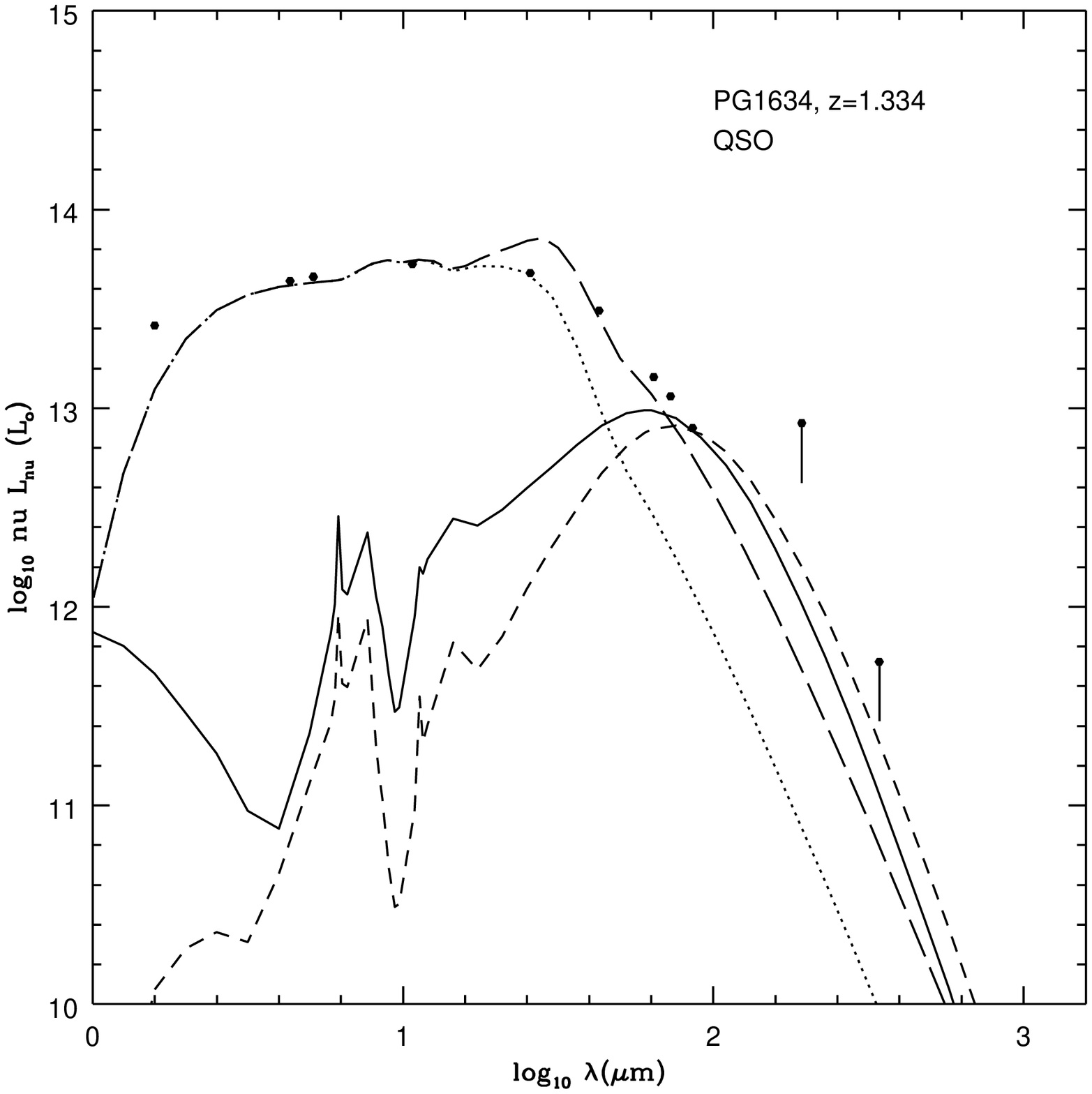,angle=0,width=8cm}
\caption{
Observed spectral energy distribution for PG1634, notation as for Fig 1.}
\end{figure}

\begin{figure}
\epsfig{file=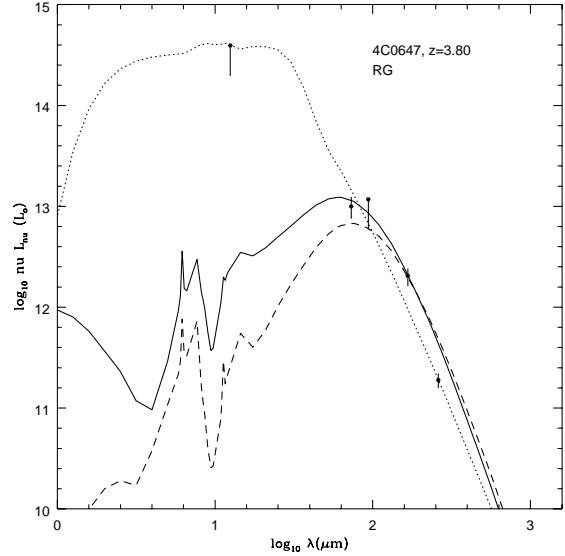,angle=0,width=8cm}
\caption{
Observed spectral energy distribution for 4C0647, notation as for Fig 1.}
\end{figure}

\begin{figure}
\epsfig{file=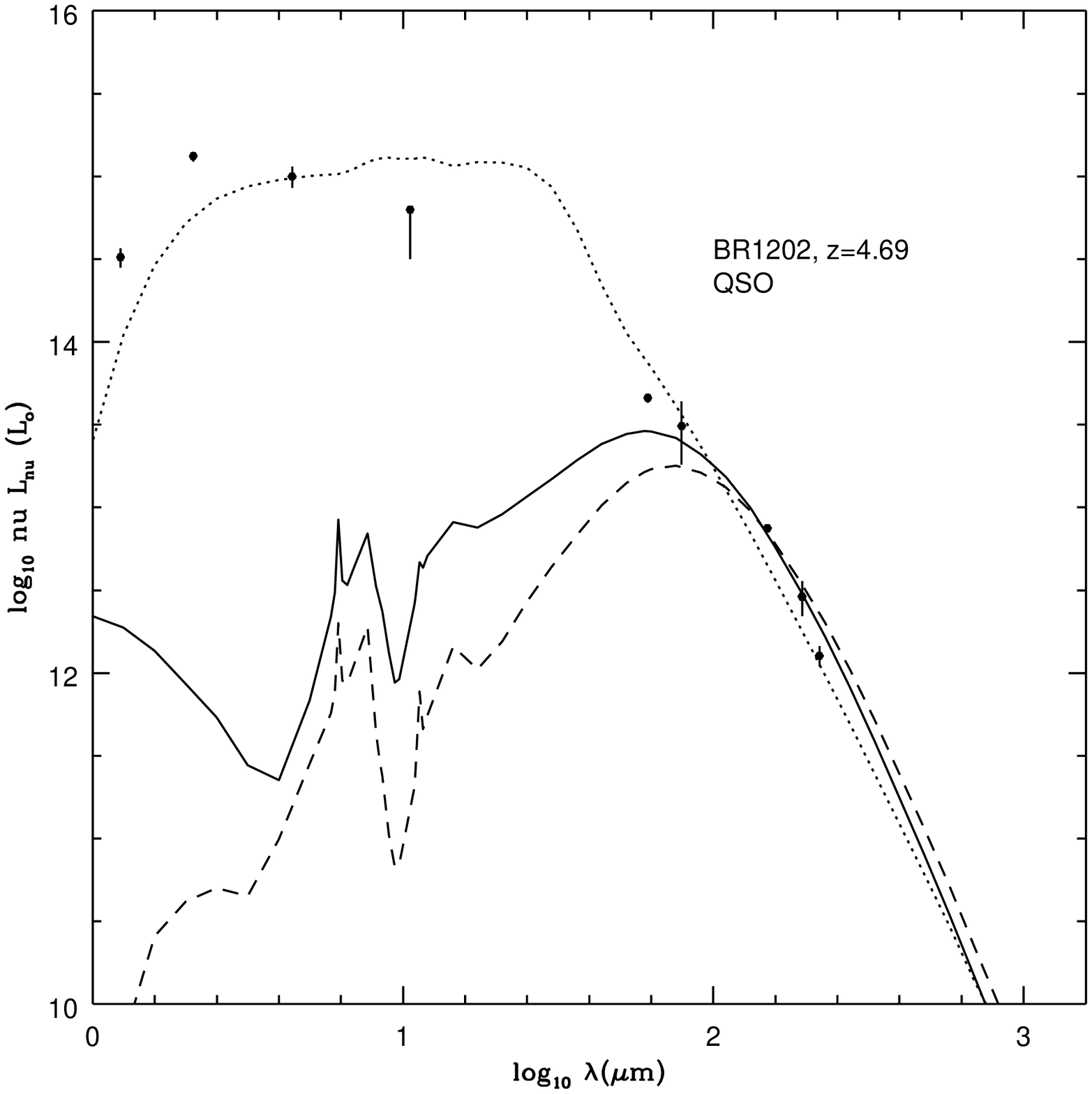,angle=0,width=8cm}
\caption{
Observed spectral energy distribution for BR1202, notation as for Fig 1.}
\end{figure}

\begin{figure}
\epsfig{file=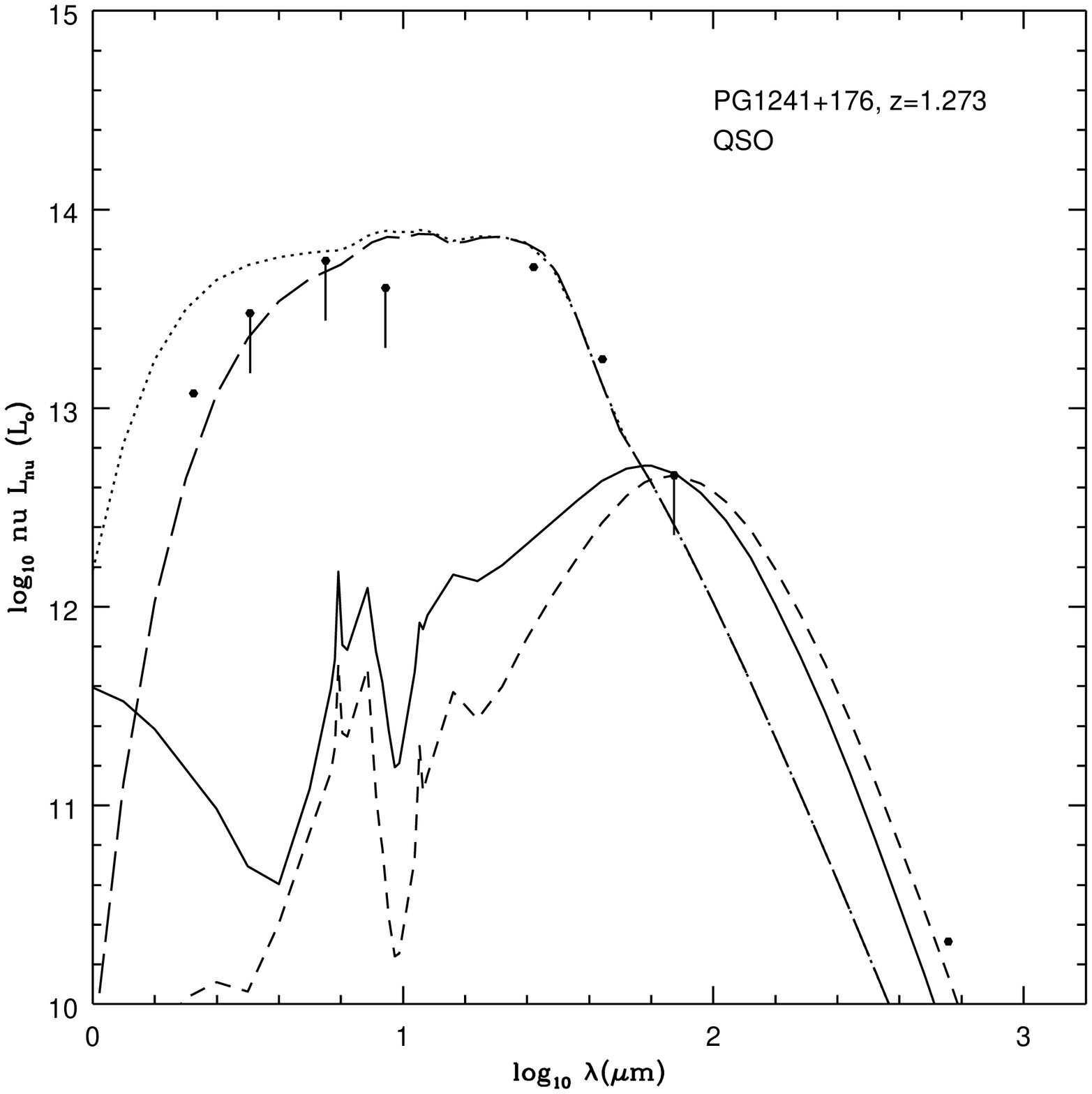,angle=0,width=8cm}
\caption{
Observed spectral energy distribution for PG1241+176, notation as for Fig 1.}
\end{figure}

\begin{figure}
\epsfig{file=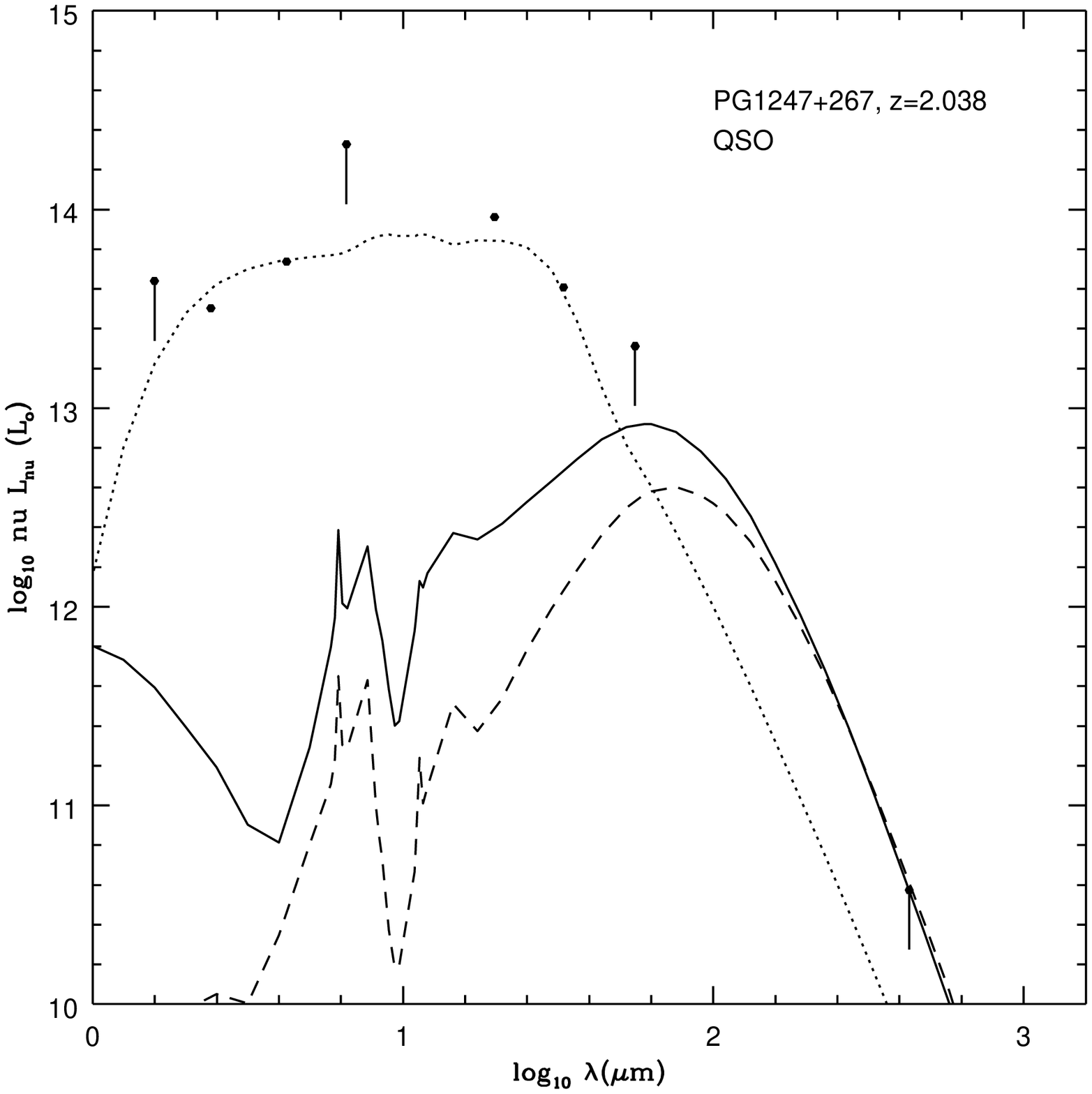,angle=0,width=8cm}
\caption{
Observed spectral energy distribution for PG1247+267, notation as for Fig 1.}
\end{figure}

\begin{figure}
\epsfig{file=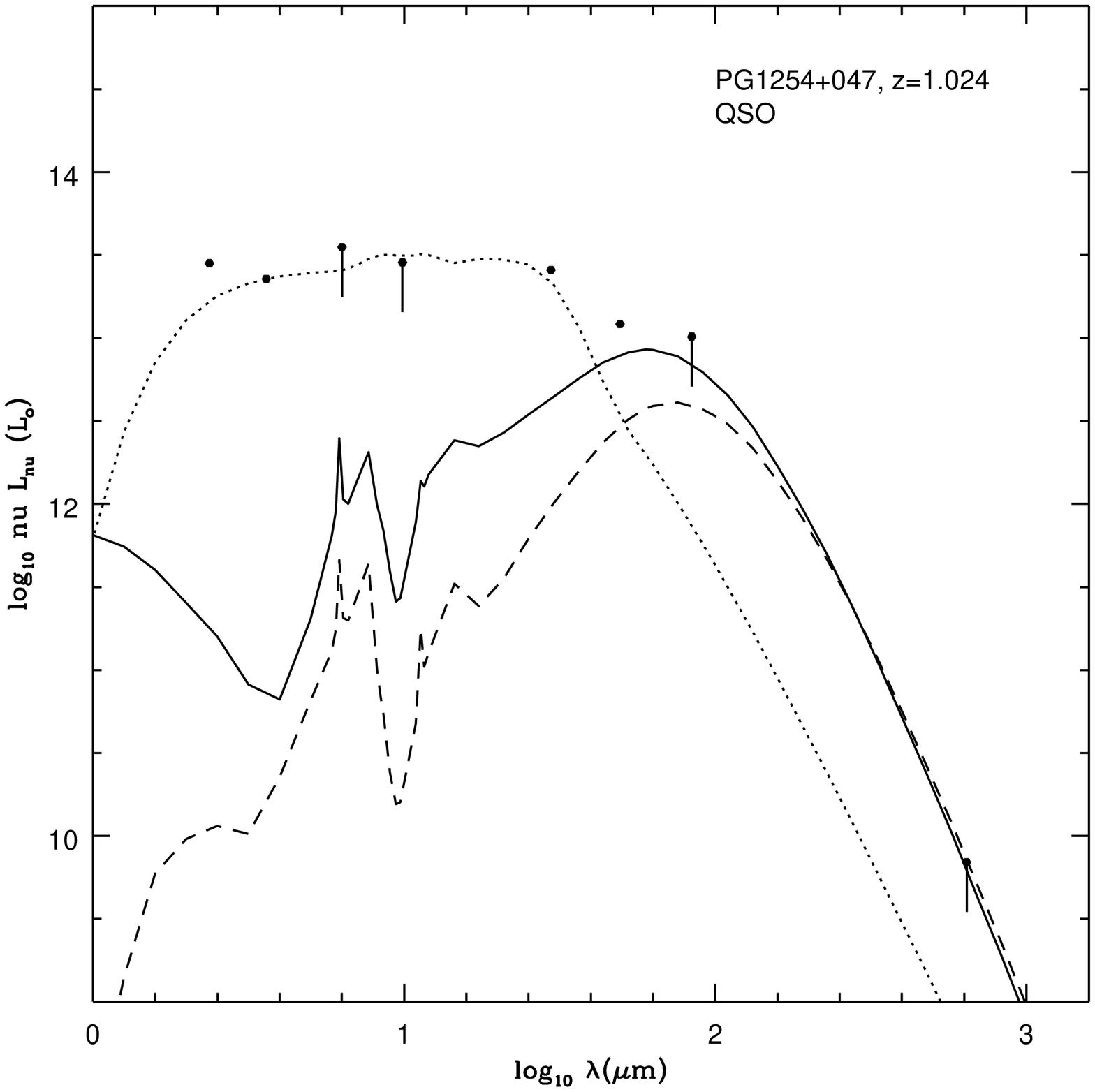,angle=0,width=8cm}
\caption{
Observed spectral energy distribution for PG1254+047, notation as for Fig 1.}
\end{figure}




On the other hand there is strong evidence for a population of galaxies with far ir 
luminosities in the 
range 1-3x$10^{13} h_{50}^{-2} L_{\odot}$ .   I have argued that in most cases the 
rest-frame radiation longward 
of 50 $\mu$m comes from a starburst component.  The luminosities are such 
as to require star formation rates in the range 3-10x$10^{3} h_{50}^{-2} M_{\odot} \/$yr, 
which would in turn generate most of the heavy elements in a $10^{11} M_{\odot}$ 
galaxy in $10^{7} -10^{8}$ yrs.  Most of 
these galaxies can therefore be considered to be undergoing their most significant 
episode of star formation, ie to be in the process of `formation'.

\section{The role of AGN}
It appears to be significant that a large fraction of these objects are Seyferts, 
radio-galaxies or QSOs.  For the 
galaxies in Tables 2 and 3, this is a selection effect in 
that 
these objects are deliberately selected to be, or to be biased towards, AGN.  
For the population of objects 
found from direct optical follow-up of IRAS samples or 850 $\mu$m surveys (Table 1), out of 12 objects, 
5 are QSOs or Seyfert 1, 
1 is Seyfert 2, and 6 are narrow-line objects.  Thus in at least 50 $\%$ of cases, 
this phase of exceptionally high far ir luminosity is accompanied by AGN activity at 
optical and uv wavelengths.  This proportion might increase if high resolution 
spectroscopy were available for all the galaxies. For comparison the proportion of 
ultraluminous galaxies which contain AGN has also been estimated as 49 $\%$ (Veilleux 
et al 1999).  However despite the high proportion
of ultraluminous and hyperluminous galaxies which contain AGN, this does not prove that an
AGN is the source of the rest-frame far infrared radiation.  The ISO-LWS mid-infrared
spectroscopic programme of
Genzel et al (1998), Lutz et al (1998), has shown that the far infrared radiation of
most ultraluminous galaxies is powered by a starburst, despite the presence of an AGN
in many cases.  Wilman et al (1999) have shown that the X-ray emission from several 
hyperluminous galaxies is too weak for them to be powered by a typical AGN.

In the Sanders et al (1989) picture, the far infrared and submillimetre emission would 
simply come from the outer 
regions of a warped disk surrounding the AGN.  Some weaknesses of this picture 
as an explanation of the far 
infrared emission from PG quasars have been highlighted by Rowan-Robinson (1995).   
A picture in which both a 
strong starburst and the AGN activity are triggered by the same interaction or merger 
event is far more likely 
to be capable of understanding all phenomena (cf Yamada 1994, Taniguchi et al 1999).

Where hyperluminous galaxies are detected at rest-frame wavelengths in the range 3-30 $\mu$m
(and this can correspond to observed wavelengths up to 150 $\mu$m), the infrared spectrum is often found
to correspond well to emission from a dust torus surrounding an AGN (eg Figs 1, 6-10, 12).  
This emission often contributes
a substantial fraction of the total infrared (1-1000 $\mu$m) bolometric luminosity. For
the 12 ir-selected objects of Table 1, the luminosity in the dust torus component
exceeds that in the starburst for 5 of the galaxies (42$\%$).  The
advocacy of this paper for luminous starbursts relate only to the rest-frame emission at 
wavelengths $\geq 50 \mu$m.  Figure 19 shows the correlation between the
luminosity in the starburst component, $L_{sb}$, and the AGN dust torus
component, $L_{tor}$, for hyperluminous infrared galaxies, PG quasars
(Rowan-Robinson 1995), and IRAS galaxies detected in all 4 bands
(Rowan-Robinson and Crawford 1989) (this extends Fig 8 of Rowan-Robinson 1995).  The range of 
the ratio between these
quantities, with 0.1 $\leq L_{sb}/L_{tor} \leq$ 10, is similar over a very wide range of infrared 
luminosity (5 orders
of magnitude), showing that the proposed separation into these
two components for hyperluminous ir galaxies is not at all implausible.

\begin{figure*}
\epsfig{file=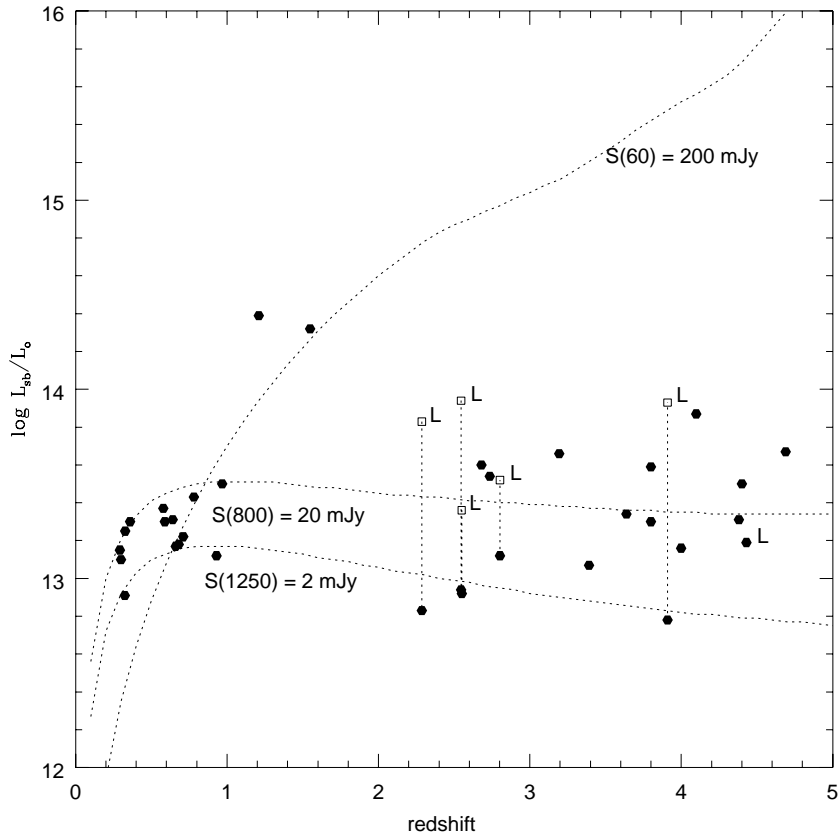,angle=0,width=12cm}
\caption{
Bolometric luminosity in starburst component for galaxies with luminosities $> 10^{13} L_{\odot}$
(Tables 1-4).  
The galaxies labelled L are known to be lensed.  Loci corresponding to the limits
set by S(60) = 200 mJy, S(800) = 20 mJy, and S(1250) = 2 mJy are shown.}
\end{figure*}

\begin{figure*}
\epsfig{file=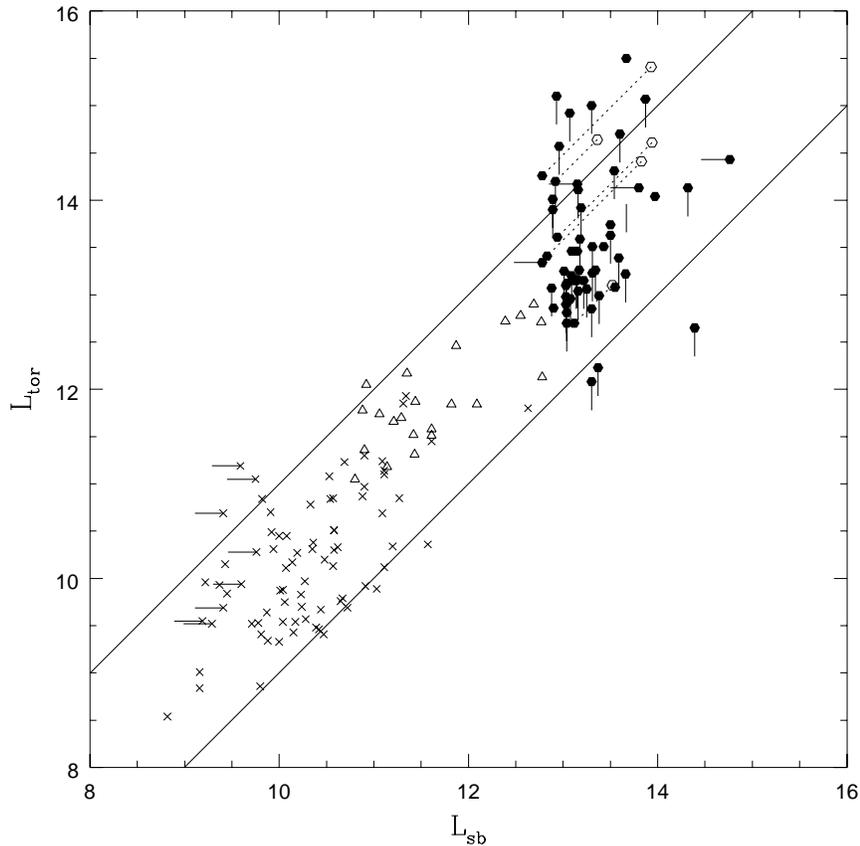,angle=0,width=12cm}
\caption{
Bolometric luminosity in  AGN dust torus component versus bolometric luminosity
in starburst component: filled circles, hyperluminous ir galaxies (this
paper); open triangles, PG quasars (Rowan-Robinson 1995); crosses, IRAS
galaxies detected in 4 bands (Rowan-Robinson and Crawford 1989, galaxies with 
only upper limits on $L_{tor}$ omitted).}
\end{figure*}

\section {Dust and gas masses}
The radiative transfer models can be used to derive dust masses and hence, via an
assumed gas-to-dust ratio, gas masses.  For the M82-like starburst model used here the
appropriate conversion is  $M_{dust} = 10^{-4.6} L_{sb}$, in solar units
(Green and Rowan-Robinson 1996).  These estimates have been converted into
estimates of gas mass assuming $M_{gas}/M_{dust}$ = 300 (tables 1-4, col 9, bracketed values).
However these estimates will not assist in deciding
the plausibility of the starburst models, because the radiative transfer
models are automatically self-consistent models of massive star-forming molecular clouds.
Estimates derived from $\nu^{\beta} B_{\nu}(T_d)$ fits to the spectral energy
distributions are even less physically illuminating.

Far more valuable are the cases where direct estimates of gas mass can be derived from
molecular line (generally CO) observations.  Where available, these estimates have been
given in Tables 1-4, col 9, taken from Frayer et al 1999a (and references therein),
Barvainis et al 1998, Evans et al (1999), Yun and Scoville (1999).  
Figure 20 shows a comparison of the
estimates of $L_{sb}$ derived here with estimates of $M_{gas}$ derived from CO observations.
Also shown are results for ultraluminous ir galaxies (Solomon et al 1997) and for
more typical IRAS galaxies (Sanders et al 1991) (after rationalisation of some objects in
common).  

The appropriate conversion factor from CO luminosity to gas mass is a
matter of some controversy.  For luminous infrared galaxies, Sanders et al (1991) 
used a characteristic value for molecular clouds in our Galaxy,  
4.78 $M_{\odot}(K km s^{-1} pc^{-2})^{-1}$.  Solomon et al (1997) found that such a value led
to gas mass estimates for ultraluminous infrared galaxies a factor of 3 or more in
excess of the dynamical masses and concluded that a value of 1.4 $M_{\odot}(K km s^{-1} pc^{-2})^{-1}$
was more appropriate for these galaxies.  Downes and Solomon (1998) studied several
ultraluminous infrared galaxies in detail in CO 2-1 and 1-0 with the IRAM interferometer,
deriving an even lower value of 0.8 $M_{\odot}(K km s^{-1} pc^{-2})^{-1}$ on the basis
of radiative transfer models for the CO lines.  However their gas masses are on average only 1/6th
of the (revised) inferred dynamical masses.  In their detailed model for Arp 220, Scoville et al (1997)
found a conversion factor 0.45 times the Galactic value, ie 2.15 $M_{\odot}(K km s^{-1} pc^{-2})^{-1}$.
Combining this with an estimated ratio for T(3-2)/T(1-0) of 0.6, Frayer et al (1999a)
justify a value of 4 $M_{\odot}(K km s^{-1} pc^{-2})^{-1}$ for gas mass estimates of
hyperluminous galaxies derived from CO 3-2 observations.

In Tables 1-4 and Fig 19, I have followed Frayer et al (1999a) in using a conversion factor
of 4 $M_{\odot}(K km s^{-1} pc^{-2})^{-1}$ for hyperluminous galaxies.  For other galaxies 
in Fig 19 with luminosities $> 10^{11.5} L_{\odot}$, I have used a conversion factor of
2 $M_{\odot}(K km s^{-1} pc^{-2})^{-1}$ (in line with Scoville et al 1997, but a factor of 2 or so
higher than advocated by Downes and Solomon 1998); for galaxies with luminosities $< 10^{11.5} L_{\odot}$,
I have used the standard Galactic value, 4.78 $M_{\odot}(K km s^{-1} pc^{-2})^{-1}$.

\begin{figure*}
\epsfig{file=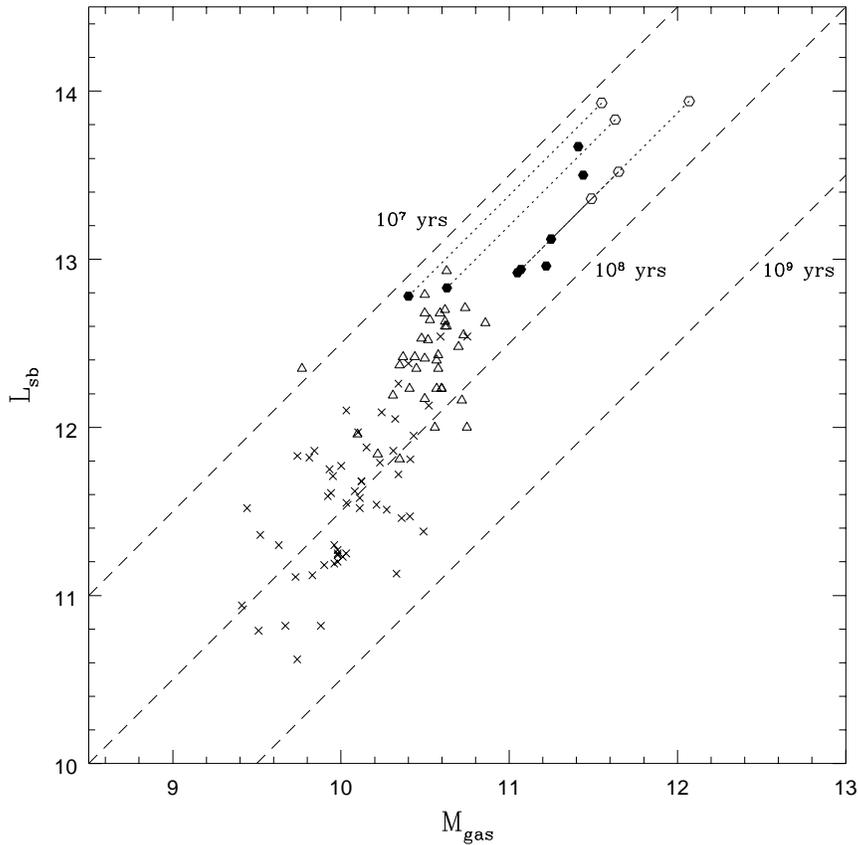,angle=0,width=12cm}
\caption{
Bolometric luminosity in starburst component, versus mass in gas, deduced
from CO observations.  Filled circles, hyperluminous ir galaxies (Frayer et al 1999,
Yun and Scovile 1999); open triangles, ultraluminous ir galaxies (Solomon et al
1997); crosses, IRAS galaxies (Sanders et al 1991).}
\end{figure*}

The range of ratios of $L_{sb}/M_{gas}$ for hyperluminous galaxies is consistent with that derived for 
ultraluminous starbursts.  For cases where we have estimates of gas mass both from
CO lines and from dust emission, the agreement is remarkably good (within a factor of 2).  
There is a tendency for the time-scale for gas-consumption, assuming a star formation
rate given by eqn (1), to be shorter for the more luminous objects, in the range
$10^7 - 10^8$ yrs (alternatively this could indicate a higher value for the low-mass
cutoff in the IMF).
The cases where a strong limit can be set on $M_{gas}$ are also, generally, those where
the seds do not support the presence of a starburst component.  After correction for the
effects of gravitational lensing, gas masses ranging up to 1-3 x $10^{11} M_{\odot}$
are seen in most hyperluminous galaxies, comparable with the total stellar mass of an $L_*$
galaxy ($10^{11.2} (M/4L) h_{50}^{-2}$).  In fact 24/39 hyperluminous galaxies in Tables 1-3
have gas masses estimated either from CO or from dust emission $> 10^{11} M_{\odot}$
(after correction for effects of lensing, where known).  
Hughes et al (1997) argue that a star-forming
galaxy can not be considered primeval unless it contains a total gas mass of 
$10^{12} M_{\odot}$, but this seems to neglect the fact that 90 $\%$ of the mass of
galaxies resides in the dark (probably non-baryonic) halo.

\section{Conclusions}

(1) About 50 $\%$ of hyperluminous infrared galaxies selected in unbiassed infrared
surveys have AGN optical spectra.  This is not in fact higher than the proportion seen in 
ultraluminous ir galaxies by Vielleux et al (199a).  For about half of the galaxies in this sample,
the AGN dust torus is the dominant contribution to the total ir (1-1000 $\mu$m) bolometric luminosity, 
while in half of cases a starburst seems to be the dominant contributor.  

(2) There is a need for both an AGN dust torus and starburst components to understand most 
seds of hyperluminous ir galaxies (29/39).  
Measured gas masses support, in most cases, the starburst interpretation of rest-frame
far-infrared and submm ($\lambda_{em} \geq 50 \mu$m)emission.  

(3) There is a broad correlation between the luminosities of starburst and AGN dust torus 
components (Fig 19).  This may imply that there is a physical link between
the triggering of star formation and the feeding of a massive black hole.  Taniguchi et al
(1999) have argued that during a merger giving rise to a luminous starburst, a pre-existing
black hole of $10^{7} M_{\odot}$ may grow into a large one $>10^{8} M_{\odot}$  and hence
form a quasar.  Alternatively they suggest that a large black hole might be formed out of star 
clusters with compact remnants. 

(3) There is no evidence in most objects that an AGN powers a significant fraction of radiation 
at rest-frame wavelengths $\geq$ 50 $\mu$m.  For P09104 and PG1634, the non-detection of CO 
emission is consistent with the absence of evidence in the sed for a starburst component.  In 
F08279, the mass of CO detected suggests a limit on the starburst luminosity which implies that 
the observed submm radiation may simply be the long-wavelength tail of its AGN dust torus emission.
F15307 poses a problem: the sed can be understood as radiation from both an AGN dust torus and
an Arp-220 like starburst, but the upper limit on the molecular mass from the non-detection
of CO would then imply a very extreme ir-luminosity to gas-mass ratio.   

(4) After correction for the effects of lensing, star-formation rates in excess of 2000 $M_{\odot}/yr$
are inferred in many of these galaxies (for a Salpeter IMF).  This would be sufficient to exhaust the observed
reservoir of gas in $10^8$ yrs.  These galaxies are undergoing extremely major episodes of star
formation, but we can not yet establish whether this is their first major burst of star formation.

(5) Further submm continuum and molecular line observations can provide a strong test of the models 
for the seds proposed here.

\begin{table*}
\caption{Hyperluminous Infrared Galaxies, found in 60 and 850 $\mu$m surveys}
\begin{tabular}{lllllllll}
name & z & $\lambda$ & flux(mJy) & ref & sp.type & lg $L_{sb}$ & lg $L_{tor}$ & lg $M_{gas}$ \\
 & & & & & & & & \\
P00182-7112 & 0.327 & 100 & 1120. & 32 & nl & & & \\
 & & 60 & 1300. & & & 13.25 (13.32) & & (11.13)\\
 & & 25 & $<$250. & & & & $<$13.06 & \\
 & & 12 & $<$250. & & & & & \\
F0023+1024 & 0.58 & 180 & 1047 $\pm$52 & 1,23 & nl & & & \\
 & & 100 & $<$938 & & &  &  & \\
 & & 90 & 923 $\pm$78 & 23 & & & &  \\
 & & 60	& 428 & & & 13.37 (13.34) & & (11.25) \\
 & & 25 & $<$193 & & & & $<$ 12.23 & \\
 & & 14.3 & $<$10.2 & 23 & & & & \\
 & & 12 & $<$173 & & & & & \\
 & & 6.7 & $<$2.45 & 23 & & & & \\
SMMJ02399-0136 & 2.803 & 1350 & 5.7 $\pm$1.0 & 25 & S2(L) & & & \\ 
 & & 850 & 26 $\pm$ 3 &  & & 13.52 (13.11) & & 11.65 (11.40) \\
 & & 750 & 28 $\pm$5 & & & & & \\
 & & 450 & 69 $\pm$15 & & & & & \\
 & & 350 & $<$323. & & & & & \\
 & & 100 & $<$715. & & & & & \\
 & & 60 & $<$428. & & & & & \\
 & & 25 & $<$86. & & & & & \\
 & & 15 & 1.2 $\pm$0.4 & & & & 13.10 & \\
 & & 12 & $<$91. & & & & & \\
P07380-2342 & 0.292 & 100 & 3550. & 32 & nl & & & \\
 & & 60 & 1170. & & & 13.09 ($\bf{13.15}$) & & (10.97) \\
 & & 25 & 800. & & & & 13.46 & \\
 & & 12 & 480. & & & & & \\
F10026+4949 & 1.12 & 100 & $<$619 & 1 & S1 & & & \\
 & & 60 & 266 & & & $<$13.84 ($<$13.90) & & \\
 & & 25 & (177) & & & & 14.04 & \\
 & & 12 & $<$86 & & & & & \\
F10214+4724 & 2.286 & 1200 & 9.6 $\pm$ 1.4 & 11 & S2(L) & & & \\
 & & 1100 & 24 $\pm$ 5 & 7 & & & & \\
 & & 800 & 50 $\pm$ 5 & & & 13.83 (13.45) & & 11.63 (11.71) \\
 & & 450 & 273 $\pm$ 45 & & & & & \\
 & & 350 & 383 $\pm$51 & 29 & & & & \\
 & & 100 & $<$510 & & & & & \\
 & & 60 & 190 $\pm$ 40 & & & & 14.41 & \\
 & & 20 & $<$45 & & & & & \\
 & & 10 & $<$12 & & & & & \\
F12509+3122 & 0.78 & 100 & $<$675 & 1 & QSO & & & \\
 & & 60 & 218 & & & 13.43 (13.63) & & (11.31)\\
 & & 25 & (103) & & & & 13.51 & \\
 & & 12 & $<$106 & & & & & \\
F1327+3401 & 0.36 & 100	& (1200) & 1 & QSO & & & \\
 & & 60 & 1180 & & & 13.30 (13.39) & & (11.18) \\
 & & 25 & $<$126 & & & & $<$ 12.85 & \\
  & & 12 & $<$94 & & & & & \\
P14026+4341 & 0.324 & 100 & (1280.) & 32 & QSO & & & \\
 & & 60 & 610. & & & 12.91 (12.99) & & (10.79) \\
 & & 25 & 260. & & & & 13.07 & \\
 & & 12 & $<$450. & & & & & \\
F14218+3845 & 1.21 & 180 & 1575 $\pm$115 & 1,23 & QSO & & & \\ 
 & & 100 & (2100) & & & & & \\
 & & 90 & 1985 $\pm$129 & 23 & & & & \\
 & & 60 & 565 & 1 & & 14.39 (14.71) & & (12.27) \\
 & & 60 & 1335 $\pm$255 & 23 & & & & \\
 & & 25	& $<$75 & & & & & \\
 & & 14.3 & $<$5.4 & 23 & & & & \\
 & & 12 & $<$97 & & & & & \\
 & & 6.7 & 1.07 $\pm$0.37 & 23 & & & $<$12.65 & \\
\end{tabular}
\end{table*}

\begin{table*}
\caption{Hyperluminous Infrared Galaxies, found in 60 and 850 $\mu$m surveys (cont.)}
\begin{tabular}{lllllllll}
name & z & $\lambda$ & flux(mJy) & ref & sp.type & lg $L_{sb}$ & lg $L_{tor}$ & lg $M_{gas}$ \\
 & & & & & & & & \\
P18216+6418 & 0.30 & 100 & 2160. & 32 & nl & & & \\
 & & 60 & 1130. & & & 13.10 (13.17) & & (10.98)\\
 & & 25 & 400. & & & & & \\
 & & 12 & (190.) & & & & 13.15 & \\
FFJ1614+3234 & 0.710 & 100 & $<$540. & 33 & nl & & & \\
 & & 60 & 174. & & & 13.22 (13.40) & & (11.10) \\
 & & 25 & $<$55. & & & & $<$13.15 & \\
 & & 12 & $<$65. & & & & & \\
F2356-0341 & 0.59 & 180 & $<$707 & 1,23 & nl & & & \\ 
 & & 100 & $<$792 & & & & & \\
 & & 90 & $<$251 & 23 & & & & \\
 & & 60 & 347 & & & 13.30 (13.45) & & (11.18) \\			
 & & 25 & $<$142 & & & & & \\
 & & 14.3 & $<$9.3 & 23 & & & & \\ 
 & & 12 & $<$87 & & & & & \\
 & & 6.7 & $<$1.74 & 23 & & & $<$12.08 & \\
\end{tabular}
\end{table*}

\begin{table*}
\caption{Hyperluminous Infrared Galaxies, found by comparison of 60 $\mu$m surveys with
quasar or radio-galaxy lists, or using an infrared colour selection biased to AGN}
\begin{tabular}{lllllllll}
name & z & $\lambda$ & flux(mJy) & ref & sp.type & lg $L_{sb}$ & lg $L_{tor}$ & lg $M_{gas}$ \\
 & & & & & & & & \\
TX0052+4710 & 1.93 & 850 & $<$9.8 & 2,22 & QSO & $<$13.15 ($<$12.73) & & ($<$11.03) \\
 & & 450. & $<$128 & 22 & & & & \\
 & & 180 & 350 $\pm$61 & 23 & & & & \\
 & & 90 & 277 $\pm$84 & 23 & & & & \\   
 & & 60	& 160 & 2 & & & 14.17 & \\
 & & 25 & $<$66 & 2 & & & & \\
 & & 14.3 & $<$9.6 & 23 & & & & \\
 & & 12 & $<$78 & 2 & & & & \\
 & & 6.7 & $<$2.49 & 23 & & & & \\
F08279+5255 & 3.91 & 1350 & 24 $\pm$2 & 26 & BALQ(L) & & & \\
 & & 850 & 75 $\pm$ 4 &  &  & $\leq$13.93 ($\leq$13.61) & & 11.55 ($<$11.81) \\
 & & 450 & 211 $\pm$47 & & & & & \\
 & & 100 & (951) & & & & & \\
 & & 60 & 511 & & & & 15.30 & \\
 & & 25 & (226) & & & & & \\
 & & 12 & $<$101 & & & & & \\
P09104+4109 & 0.44 & 100 & $<$438 & 6 & S2 & & & \\
 & & 60 & 525 & & & $<$12.78 ($<$12.70) & & $<$10.48 ($<$10.66) \\
 & & 25 & 333 & & & & & \\
 & & 12 & (130) & & & & 13.39 & \\
TX1011+1430 & 1.55 & 100 & $<$757 & 2 & QSO & & & \\
 & & 60 & 225 & & & 14.32 (14.72) & $\leq$ 14.13 & (12.20) \\
 & &  25 & $<$269 & & & & \\
 & & 12 & $<$116 & & & & \\
PG1148+549 & 0.969 & 100 & 410 & 9 & QSO & 13.50 (13.53) & & (11.38) \\
 & & 60 & 196 & 9 & & & & \\
 & & 25 & 120 & 9 & & & 13.74 & \\
 & & 12 & $<$75 & 9 & & & & \\
PG1206+459 & 1.158 & 170 & 188 & 36 & QSO & & & \\
 & & 100 & 386 & 36 &  & $<$13.13 ($<$13.00) & & ($<$11.01) \\
 & & 60 & 463 & 36 & & & & \\
 & & 25 & $<$113 & 9 & & & & \\
 & & 20 & 208 & 36 & & & 14.21 & \\
 & & 12.8 & 110 & 36 & & & & \\
 & & 12 & 207 $\pm$36 & 9 & & & & \\
 & & 4.8 & $<$10 & 36 & & & & \\
PG1248+401 & 1.030 & 100 & $<$378 & 9 & QSO & $<$13.53 ($<$13.57) & & ($<$11.41)\\
 & & 60 & 224 $\pm$51 & 9 & & & 13.83 & \\
 & & 25 & $<$ 200 & 9 & & & & \\
 & & 12 & $<$117 & 9 & & & & \\
H1413+117 & 2.546 & 1250 & 18 $\pm$ 2 & 21 & BALQ(L) & & & \\
 & & 800 & 66 $\pm$ 7 & 18 & & 13.94 (13.54) & & 12.07 (11.82) \\	
 & & 761 & 44 $\pm$ 8 & 12 & & & & \\	
 & & 438 & 224 $\pm$ 38 & 12 & & & & \\
 & & 345 & 189 $\pm$ 56 & 12 & & & & \\
 & & 350 & 293 $\pm$14 & 29 & & & & \\
 & & 200 & 280 $\pm$50 & 36 & & & & \\ 
 & & 100 & 370 $\pm$ 78 & 21 & & & & \\	
 & & 60 & 230 $\pm$ 38 & 21 & & & 14.61 & \\
 & & 25 & 75 $\pm$3 & 36 & & & & \\
 & & 12	& 57 $\pm$8 & 36 & & & & \\
 & & 7 & 17 $\pm$2 & 36 & & & & \\
 & & 4.8 & 10 $\pm$3 & 36 & & & & \\
F14481+4454 & 0.66 & 100 & $<$500 & 27 & S2 & & & \\
  & & 60 & 190 & & & 13.17 (13.34) & & (11.05) \\
  & & 25 & (85) & & & & 13.26 & \\
  & & 12 & $<$76 & & & & & \\
F14537+1950 & 0.64 & 100 & $<$738 & 27 & sb & & & \\
 & & 60 & 283 & & & 13.31 (13.46) & & (11.19) \\
 & & 25 & $<$ 159 & & & & $<$ 13.51 & \\
 & & 12 & $<$ 119 & & & & & \\
\end{tabular}
\end{table*}
 
\begin{table*}
\caption{Hyperluminous Infrared Galaxies, found by comparison of 60 $\mu$m surveys with
quasar or radio-galaxy lists, or using an infrared colour selection biased to AGN}
\begin{tabular}{lllllllll}
name & z & $\lambda$ & flux(mJy) & ref & sp.type & lg $L_{sb}$ & lg $L_{tor}$ & lg $M_{gas}$ \\
 & & & & & & & & \\
F15307+325 & 0.93 & 3000. & $<$1.3 & 31 & S2 & & & \\
 & & 1250 & $<$5.1 & 31 & & & & \\
 & & 180 & 397 $\pm$67 & 23 & & & & \\   
 & & 100 & 510 $\pm$ 62 & 16 & & 13.55 ($\bf{13.12}$)& & $<$10.39 (11.43) \\
 & & 90 & 478 $\pm$65 & 23 & & & & \\
 & & 60 & 280 $\pm$ 42 & & & & & \\
 & & 25 & 80 $\pm$ 24 & & & & 13.08 & \\
 & & 14.3 & 15.3 $\pm$2.6 & 23 & & & & \\
 & & 12 & $<$45 & & & & & \\
 & & 6.7 & 2.95 $\pm$0.51 & 23 & & & & \\
PG1634+706 & 1.334 & 800 & $<$47 & 17 & QSO & & & \\
 & & 450 & $<$420 & 17 & & & & \\
 & & 200 & 216 & 35 & & & & \\
 & & 170 & 238 & 35 & & & & \\
 & & 150 & 307 & 35 & & & & \\ 
 & & 100 & 343 & 9  & & $<$13.20 ($<$13.05) & & $<$10.70 ($<$11.08)\\
 & & 60 & 318 & 9 & & & 14.13 & \\
 & & 25 & 147 & 9 & & & & \\
 & & 12 & 61 & 9 & & & & \\
\end{tabular}
\end{table*}

\begin{table*}
\caption{Hyperluminous Infrared Galaxies, found through submm observations of known high redshift AGN}
\begin{tabular}{lllllllll}
name & z & $\lambda$ & flux(mJy) & ref & sp.type & lg $L_{sb}$ & lg $L_{tor}$ & lg $M_{gas}$ \\
 & & & & & & & & \\
HM0000-263 & 4.10 & 350 & 134 $\pm$29 & 29 & QSO & 13.87 (13.79) & $<$15.07 & (11.75) \\
Q0100+1300 &  2.68 & 350 & 131 $\pm$28 & 29 & QSO & 13.60 (13.42) & $<$ 14.70 & (11.48) \\
PC0307+0222 & 4.379 & 1250 & 6.6 $\pm$ 1.7 & 4 & QSO & 13.31 (12.92) & $<$13.23 & (11.19) \\
 & & R & 20.39 & 20 & & & & \\
PC0345+0135 & 3.638 & 1250 & 6.1 $\pm$ 2.0 & 4 & QSO & 13.34 (12.93) & $<$13.26 & (11.22) \\
 & & 800 & $<$25 & 18 & & & & \\
 & & R & 19.94 & 20 & & & & \\
MG0414+0534 & 2.639 & 350 & $<$105. & 29,34 & QSO(L) & & & \\
 & & 60 & 140. $\pm$40. & & & $<$14.76 ($<$15.3) & 14.43 & 11.53 ($<$12.64)\\
 & & 25 & 70. $\pm$24. & & & & & \\
4C0647+4134 & 3.8 & 1250 & 2.5 $\pm$ 0.4& 5 & RG & & & \\
 (4C41.17)& & 800 & 17.4 $\pm$ 3.1 & 5 & & 13.30 (12.97) & $<$15.00 & (11.18) \\
 & & 450 & $<$56 & 5 & & & & \\
 & & 350 & 37 $\pm$9 & 29 & & & \\
BR1202-0725 & 4.69 & 1250 & 10.5 $\pm$ 1.5 & 18 & QSO & $\leq$13.67 ($\leq$13.39) & 15.50 & 11.41 ($<$11.55)\\
 & & 1100 & 21 $\pm$ 5 & 8 & & & & \\
 & & 850 & 42 $\pm$ 2 & 8 & & & & \\
 & & 450 & 92 $\pm$ 38 & 8 & & & & \\
 & & 350 & 106 $\pm$ 7 & 29 & & & & \\
 & & 25 & 165 $\pm$25 & 38 & & & & \\
 & & 12 & 105 $\pm$7 & 38 & & & & \\
 & & 7 & 15 $\pm$2 & 38 & & & & \\
 & & 4 & $<$6 & 38 & & & & \\
 & & R & 18.7 & 20 & & & & \\
PG1241+176 & 1.273 & 1300 & 3.3 & 37 & QSO & & & \\
 & & 170 & $<$96 & & & $<$12.92 ($<$12.80) & & ($<$10.80) \\
 & & 100 & 217 & & & & & \\
 & & 60 & 378 & & & & 14.28 & \\
 & & 12.8 & $<$87 & & & & & \\
 & & 7.3 & $<$27 & & & & & \\
 & & 4.8 & 7 & & & & & \\ 
PG1247+267 & 2.038 & 1300 & $<$2.1 & 37 & QSO & $<$13.13 ($<$12.74) & & ($<$11.01) \\
 & & 170 & $<$150 & & & & & \\
 & & 100 & 174 & & & & & \\
 & & 60 & 236 & & & & 14.26 & \\
 & & 12.8 & 30 & & & & & \\
 & & 7.3 & 10 & & & & & \\
 & & 4.8 & $<$9 & & & & & \\ 
PG1254+047 & 1.024 & 1300 & $<$1.8 & 37 & QSO & $\leq$13.14 ($\leq$12.75) & & ($<$11.02) \\
 & & 170 & $<$345 & & & & & \\
 & & 100 & 242 & & & & & \\
 & & 60 & 307 & & & & 14.28 & \\
 & & 12.8 & $<$90 & & & & & \\
 & & 7.3 & 33 & & & & & \\
 & & 4.8 & 27 & & & & & \\ 
LBQS1230+1627 & 2.735 & 1250 & 7.5 $\pm$ 1.4 & 19 & QSO & 13.54 (13.05) & $<$ 14.31 & (11.42) \\
 & & 350 & 104 $\pm$21 & 29 & & & & \\
 & & V & 17.4 & 30 & & & & \\
BRI1335-0417 & 4.396 & 1250 & 10.3 $\pm$ 1.04 & 19 & QSO & 13.50 (13.12) & $<$13.63 & 11.44 (11.38) \\
 & & 850 & 14 $\pm$ 1 & 24 & & & & \\
 & & 350 & 52 $\pm$ 8 & 29 & & & & \\
 & & R & 19.4 & 20 & & & & \\
PC2047+0123 & 3.799 & 1250 & 2.08 $\pm$ 0.47 & 13 & QSO & 13.59 (13.49) & $<$13.39 & (11.47) \\
 & & 350 & 80 $\pm$20 & 29 & & & & \\
 & & R & 19.71 & 30 & & & & \\
PC2132+0126 & 3.194 & 1250 & 11.5 $\pm$ 1.7 & 4 & QSO & 13.66 (13.19) & $<$13.22 & (11.54) \\
 & & 800 & $<$12 & 18 & & & & \\
 & & R & 19.78 & 30 & & & & \\
\end{tabular}
\end{table*}
\medskip

\begin{table*}
\begin{minipage}{140mm}
\caption{Luminous Infrared Galaxies not meeting the requirements for Tables 1-3, but
satisfying $L_{sb} + L_{tor} > 10^{13.0}$.}
\begin{tabular}{lllllllll}
name & z & $\lambda$ & flux(mJy) & ref & sp.type & lg $L_{sb}$ & lg $L_{tor}$ & lg $M_{gas}$ \\
 & & & & & & & & \\
BR0902+34 & 3.391 & 1250 & 3.1 $\pm$0.6 & 18 & RG & 13.07 & $<$14.92 & (10.95) \\
  & & 850 & $<$14 & & & & & \\
  & & 450 & $<$99 & & & & & \\
BRI0952-0115 & 4.43 & 1250 & 2.78 $\pm$ 0.63 & 19 & QSO(L) & & & \\
 & & 850 & 14 $\pm$ 2 & 24 & & 13.19 (12.91) & $<$13.92 & (11.07) \\
 & & 350 & $<$66 & 29 & & & & \\
 & & R & 18.7 & 20 & & & & \\
BR1117-1329 & 4.00 & 1250 & 4.09 $\pm$ 0.81 & 19 & QSO & 13.16 (12.86) & $<$14.11 & (11.04) \\
 & & 850 & 13 $\pm$ 1 & 24 & & & & \\
 & & 350 & $<$39 & 29 & & & & \\
 & & R & 18.0 & 20 & & & & \\
F1220+0939 & 0.68 & 60 & 180 & 1 & QSO & 13.18 (13.36) & $\leq$13.59 & (11.06) \\
F12514+1027 & 0.30 & 100 & 755. & 27 & S2 & & & \\
 & & 60 & 712. & & & 12.90 (12.97) & & (10.78) \\
 & & 25 & 190. & & & & 12.86 & \\
 & & 12 & $<$63.2 & & & & & \\
SMMJ14011+0252 & 2.55 & 1350 & 6.06 $\pm$1.46 & 28 & nl(L) & & & \\
 & & 850 & 14.6 $\pm$1.8 & & & 13.36 (12.97) & $<$ 14.64 & 11.49 (11.24) \\
 & & 450 & 41.9 $\pm$6.9 & & & & & \\
\end{tabular}
\medskip 

1	McMahon, Rowan-Robinson et al 1999,
2	Dey and van Breugel 1995,
3	van Ojik et al 1994,
4	Andreani et al 1993,
5	Dunlop et al 1995,
6	Kleinmann et al 1988,
7	Rowan-Robinson et al 1991, 1993,
8	Isaak et al 1994,
9	Sanders et al 1989, Rowan-Robinson 1995,
10 	McMahon et al 1995,
11	Downes et al 1992,
12	Barvainis et al 1992,
13	Ivison 1995,
16	Cutri et al 1994,
17	Hughes et al 1993,
18	Hughes et al 1997,
19	Omont et al 1996,
20	Storrie-Lombard et al 1996,
21	Barvainis et al 1995,
22	Ivison 1999 (personal comm.),
23	Verma et al 1999,
24	McMahon et al 1999,
25	Ivison et al 1998,
26	Irwin et al 1999, Lewis et al 1999,
27 	Cutri et al 1999 (in prep.), Wilman et al 1999
28	Ivison et al 1999,
29	Benford et al 1999,
30	Veron et al quasar catalogue
31  	Yun and Scoville 1999,
32	Saunders et al 1996
33	Tran et al 1999,
34	Barvainis et al 1998,
35 	Haas et al 1998
36	ISO data archive
37 	Haas et al 1999
38	Wilkes 1999 (personal communication)
\end{minipage}
\end{table*}

\end{document}